\documentclass[a4paper,11pt]{article}

\usepackage{graphicx}
\usepackage{caption}
\usepackage{subcaption}
\usepackage{slashed,nicefrac}

\usepackage{mathtools}

\usepackage{fullpage}
\usepackage[T1]{fontenc}

\usepackage{amssymb,amsmath,cite}

\usepackage{xcolor,bbold,comment}

\usepackage{amsthm}

\begin{document}
\vspace{5mm}
\vspace{0.5cm}

\def\thefootnote{\arabic{footnote}}
\setcounter{footnote}{0}

\allowdisplaybreaks 

\begin{titlepage}
\thispagestyle{empty}

\begin{flushright}
	\hfill{\ } 
\end{flushright}
				
\vspace{35pt}
				
\begin{center} 
	{\LARGE{\bf 
	The unbearable lightness of charged gravitini 
	}} 
									
	\vspace{50pt}
							
	{Gianguido~Dall'Agata, Maxim~Emelin, Fotis~Farakos and Matteo~Morittu}
							
	\vspace{25pt}
							
	{

		{\it Dipartimento di Fisica e Astronomia ``Galileo Galilei''\\
			Universit\`a di Padova, Via Marzolo 8, 35131 Padova, Italy}
										
		\vspace{15pt}
										
		{\it  INFN, Sezione di Padova \\
		Via Marzolo 8, 35131 Padova, Italy}
		}
								
\vspace{40pt}
								
{ABSTRACT} 
\end{center}

We prove that charged gravitini cannot have parametrically small or vanishing Lagrangian mass 
in de Sitter vacua of extended supergravity while respecting the magnetic weak gravity conjecture. 
This places large classes of de Sitter solutions of gauged supergravity in the swampland, including all known stable solutions of the N=2 theory. 
We illustrate this result by analyzing a variety of de Sitter critical points of N=2 matter-coupled supergravity that also include new stable de Sitter solutions. 
Our results provide concrete evidence that (quasi) de Sitter with charged light gravitini should belong to the swampland, which also strongly resonates with the ``festina lente'' bound.

\vspace{10pt}
			
\bigskip
			
\end{titlepage}

\numberwithin{equation}{section}

\baselineskip 6.006 mm

\tableofcontents

\newpage

\section{Introduction}

The vacuum structure of four-dimensional gauged supergravities has been under intense scrutiny ever since the first matter-coupled theories were constructed and their close connection to string theory was appreciated. 
However, it is not always evident which supergravity models are true low-energy effective theories arising from string compactifications, which are truncations of larger low-energy field theories, and which may have no relation to string theory at all. 
To make progress on this question one can turn to the swampland criteria \cite{Palti:2019mpp}, a set of interconnected conjectures that aim to restrict the properties of low-energy effective field theories (EFTs) that have a consistent embedding into quantum gravity.

The swampland criteria are motivated by black hole arguments, quantum gravity arguments, or simply by recurring patterns in string flux compactifications. 
Arguably, among the most well-established of the swampland criteria is the weak gravity conjecture (WGC) \cite{Arkani-Hamed:2006emk}. 
The ``electric'' version of the WGC places restrictions on the possible masses and charges that can exist in an EFT, 
whereas the ``magnetic'' version further posits that there is a prematurely low ultraviolet (UV) cut-off that depends on the gauge coupling. 
In parallel, various considerations have cast doubt on the possibility of realizing de Sitter space within string theory \cite{Danielsson:2018ztv}, or in quantum gravity more broadly \cite{Dvali:2018jhn}. 
This culminated in a series of conjectures that explicitly forbid stable \cite{Obied:2018sgi, Andriot:2018wzk, Garg:2018reu, Ooguri:2018wrx} or long-lived \cite{Bedroya:2019snp, Bedroya:2019tba, Bedroya:2020rac} de Sitter solutions in a UV-completable EFT. 
These conjectures, if correct, have profound implications for string cosmology. 
It would be therefore desirable to acquire further evidence for or against these conjectures, in particular by exploring their connections to other more established ones. 
In this vein, it was demonstrated in \cite{Cribiori:2020use} that applying the magnetic WGC on certain supergravity de Sitter vacua results in their elimination without the need to invoke other conjectures. 
It was further noted that this violation of the WGC correlates with the presence of charged massless gravitini which was proposed as an independent swampland criterion in its own right. 

In this work we further build on the results of \cite{Cribiori:2020use} using N=2 matter-coupled gauged supergravity as our main framework  \cite{Lauria:2020rhc,DallAgataxxx}. 
We present for the first time a general proof that parametrically light or massless\footnote{We stress that the mass is not a good quantum number to describe physical states in de Sitter and Anti-de Sitter spacetimes, because $P^2$ is not a good Casimir of the corresponding symmetry algebras. 
Also,  gravitini are never truly massless in de Sitter, even when they have vanishing Lagrangian mass,  in the sense that they always propagate four degrees of freedom instead of two. 
Therefore in our work when we refer to ``massless gravitini'' we mean gravitini with vanishing Lagrangian mass, whereas by ``light gravitini'' we will mean a vanishing or parametrically small Lagrangian mass compared to the Hubble scale. 
We will omit the word ``Lagrangian'' to avoid clutter and we hope the reader will not be confused by this omission.} 
charged gravitini at a de Sitter critical point result in a violation of the magnetic WGC in N=2 supergravity. 
This result is illustrated with several N=2 models that have de Sitter critical points, both stable or unstable, or even have flat directions corresponding to moduli. 
We are able to exclude a majority of these de Sitter solutions, including some that fail to be excluded even by the de Sitter swampland criteria. 
We also discuss examples that evade exclusion by the WGC by having either uncharged gravitini or by breaking all gauge symmetry at the critical point. 
After examining the N=2 models, we also provide a parallel proof that de Sitter critical points with light charged gravitini are similarly excluded in N=8 supergravity. 
Our findings strongly indicate that an analogous result for other extended supergravities with 8>N>2 should hold. 

As a cross-validation of our findings, we see that our results are strongly consonant with the ``festina lente'' (FL) bound \cite{Montero:2019ekk,Montero:2020rpl,Montero:2021otb}, which places a lower bound on the masses of all charged particles in a de Sitter background. 
Conversely, our findings can be interpreted as a highly non-trivial check on the consistency of the FL bound, by simply applying it on the N=2 gravitini and demanding gravity to be the weakest force. 
Complementary arguments, that deal with the lowering of the EFT cut-off in the limit of light gravitini, including considerations based on the WGC, appear in \cite{Cribiori:2021gbf}, and are further established in \cite{Castellano:2021yye}.\footnote{For a different line of investigation concerning massless gravitini based on post-inflationary cosmological considerations, see \cite{Kolb:2021xfn,Dudas:2021njv,Terada:2021rtp,Antoniadis:2021jtg}.} 
For completeness, let us note that further restrictions on the properties of scalar potentials such that gravity remains the weakest force have been considered in \cite{Palti:2017elp,Gonzalo:2019gjp,Dall'Agata:2020}, while the WGC in de Sitter space has been further studied in \cite{Huang:2006hc,Antoniadis:2020xso}.

\section{General considerations}
\label{general}

In this section we present our main arguments for why (quasi) de Sitter backgrounds with charged light gravitini belong to the swampland. 
First we recall how the magnetic weak gravity conjecture for a U(1) places a restriction on the energy density of a theory and then we further argue for an extension of this restriction to the case of non-abelian gauge symmetry. 
After that we present a general proof that de Sitter critical points in N=2 gauged supergravity with charged massless gravitini belong to the swampland and we show that it also applies to the case of parametrically light masses. 
We close the section by relating our results to other swampland conjectures. 

\subsection{Magnetic WGC and de Sitter} 

In this section we review the implications that the weak gravity conjecture can have on de Sitter backgrounds. 
Let us first recall that the magnetic WGC postulates that for any U(1) gauge symmetry there is a quantum gravity-induced UV cut-off \cite{Arkani-Hamed:2006emk}. 
The value of that cut-off $\Lambda_{UV}$ for an EFT is bounded from above by the formula 
\begin{align}
\label{mwgc}
\Lambda_{UV} < g_{U(1)} \, q_{el} \, M_P \ , \qquad \text{for every charged object} \,, 
\end{align} 
where $g_{U(1)}$ is the U(1) gauge coupling and $q_{el}$ is the charge with respect to that U(1). 
From now on we will call $g_{U(1)} q_{el}$ the physical coupling of an object 
\begin{align}
q_{phys} = g_{U(1)} q_{el} \,. 
\end{align}
Clearly the object with the lowest physical coupling sets the strongest restriction on the allowed UV cut-off, and for uncharged objects \eqref{mwgc} does not apply. 
The way the UV cut-off manifests in the EFT is not known a priori unless one knows also the UV completion of the theory. 
For example it can be due to towers of massive states that have a mass controlled by $\Lambda_{UV}$. 
For us here this UV cut-off will be simply used as a device to signal when higher order corrections to the effective theory become important. 
If one wants to safely ignore such corrections, then one should work at energies parametrically lower than $\Lambda_{UV}$. 

On a de Sitter background there is one simple condition that should be satisfied such that higher order gravitational corrections do not immediately threaten the EFT. 
For a background with Hubble constant $H$ the condition is 
\begin{align}
\label{HllUV}
H \ll \Lambda_{UV} \,. 
\end{align}
An extended discussion justifying this condition can be found in \cite{Cribiori:2020use}. 
If \eqref{HllUV} does not hold then the two-derivative gravitational theory may be subject to strong quantum corrections. As a result it is not a trustworthy EFT.

\subsubsection{Warm-up: Gauged R-symmetry in N=1} 

A simple illustration of the restrictions placed by the magnetic WGC on supergravity theories is the following observation that was presented in \cite{Cribiori:2020wch}. 
If we consider the Freedman model \cite{Freedman:1976uk} then the Lagrangian contains only gravitation with a positive cosmological constant, 
a U(1) gauge field ($v_\mu$) that gauges the R-symmetry, and a massless, but charged, gravitino ($\psi_\mu$). 
In the unitary gauge this is
\begin{equation}
\label{freedman}
\begin{aligned}
e^{-1} {\cal L} 
= &   -\frac{1}{2} R 
+ \frac12 \epsilon^{\kappa \lambda \mu \nu} \left( \overline \psi_\kappa \overline \sigma_\lambda D_\mu \psi_\nu 
- \psi_\kappa \sigma_\lambda D_\mu \overline \psi_\nu  \right)  \\[2mm]
& -\frac{1}{4g^2} F_{\mu\nu} F^{\mu\nu} + i  q \, \epsilon^{\kappa \lambda \mu \nu} \overline \psi_\kappa \overline \sigma_\lambda \psi_\mu v_\nu -  4 g^2 q^2  \, , 
\end{aligned}
\end{equation}
where $D_\mu$ here is the spacetime Lorentz-covariant derivative, which includes the gravitino-dependent spin-connection. 
This de Sitter space is characterized by a Hubble scale that is of the same order as the gravitino charge times the gauge coupling. 
As a result one can argue that the Hubble scale of such a simple model already hits the magnetic WGC cut-off and thus is faced with a Dine--Seiberg problem \cite{Dine:1985he}. 
Interestingly, one could place this model in the swampland equally well just by applying the FL conjecture \cite{Montero:2021otb}. 

\subsection{WGC and non-abelian gauge groups} \label{non-abelian}

The magnetic WGC is formulated for U(1) gauge symmetries, and a natural question is whether a similar expression for the UV cut-off exists involving the gauge couplings of non-abelian groups. 
In theories with charged scalar fields one should expect this to be the case, because the gauge group itself can be broken or enhanced depending on the expectation values of the fields.
This is especially true in gauged extended supergravities, where the gauging of a non-abelian group forces the theory to contain the would-be goldstones of the non-abelian gauge symmetries.

The simplest case to consider would be a theory where a certain vacuum preserves a non-abelian gauge symmetry but also has a charged modulus, such that giving it an expectation value breaks the non-abelian symmetry to a U(1). 
For any value of this modulus, we have a vacuum of the theory, where one can clearly apply the WGC using the gauge coupling of this U(1). 
We can then take the limit as the expectation value approaches the original ``central'' vacuum where the full non-abelian symmetry is restored. 
In this limit the U(1) gauge coupling will approach the non-abelian coupling of the central vacuum. 

In this sequence of U(1) gauge theories we can determine the WGC cut-off in the usual manner and then by continuity, we conclude that the cut-off of the vacuum with non-abelian symmetry must be the limit of the cut-offs of the broken phase. 
Since this cut-off is determined from the U(1) gauge coupling, which in turn approaches the non-abelian coupling, we conclude that the non-abelian coupling can also be used to determine the UV cut-off for the original non-abelian theory. 

If we interpret the WGC cut-off as, for example, coming from the mass scale of some UV states that aren't captured by the effective theory, these masses would not change drastically between nearby points in moduli space. 
We expect this to be the case, regardless of any gauge symmetry enhancement or breaking that may also be taking place. 
Thus if the WGC is a good criterion for determining the cut-off of a U(1) gauge theory, it should work when we approach a point where this symmetry is enhanced. 
The same holds for any other UV origin of the cut-off.

A more subtle situation happens when the direction in field space that breaks the non-abelian symmetry but preserves a U(1) is not a modulus. 
In this case, although we can still consider a sequence of points in moduli space that approach the central vacuum, they will not dynamically ``stay in place'', so to speak. 

In that case, the next best scenario is if the U(1) preserving direction also has the gradient of the potential tangent to it. 
Then there exists a U(1)-preserving classical trajectory, and the situation is similar to the case when it's a modulus. 
Indeed, if we set conditions at $t = 0$ where this field has a sufficiently small but non-zero expectation value and vanishing kinetic energy, the subsequent trajectory will be along the U(1)-preserving direction. 
In this case we can expand the action around that path and this expansion will still have a massless U(1) gauge field manifestly present, with the various field excitations charged under it. 
The magnetic WGC can then be applied as usual to determine a (possibly time-dependent) UV cut-off for the effective theory defined by the expansion around such a non-stationary classical path.

If we can carry out this procedure of defining effective theories around U(1)-preserving non-stationary backgrounds, we can consider a sequence of such effective theories defined around paths with $t=0$ conditions closer and closer to the central vacuum. 
Once again we expect the cut-offs for this sequence of effective theories to approach the cut-off of the non-abelian theory, while the U(1) gauge coupling will approach the non-abelian coupling, leading to the non-abelian WGC.

Of course in general theories, we do not always have the ability to break the non-abelian group while preserving a U(1) subgroup.
The above arguments are particularly relevant to N=2 supergravity, where the vector multiplet scalars transform in the adjoint of the non-abelian gauge group and can thus be used to break it to a U(1). 
In section \ref{so3} we will see examples of both of the above scenarios, where an SU(2) gauge group in the central vacuum gets broken down to a U(1) either by a modulus or a tachyonic scalar that allows for U(1)-preserving classical trajectories. 
In both of those examples we will be able to exclude the central vacuum with non-abelian gauge group by considering the WGC for the neighboring U(1)-preserving points. 

A final possible caveat is that although a U(1) preserving direction in field space might exist, there might be no U(1) preserving classical trajectories, if the gradient of the potential is not aligned with the U(1) preserving direction. 
In this case it isn't clear how to apply the WGC. 
We have not encountered such examples in our investigations.

\subsection{N=2 with charged light gravitini}\label{chargedmassless} 

Here we provide a simple proof of the fact that in N=2 gauged supergravity de Sitter critical points with charged massless gravitini are incompatible with the consistency requirements of the weak gravity conjecture. 
For the sake of clarity in the presentation we directly give the argument in the following, using only the N=2 ingredients that are directly relevant. 
The interested reader can find a summary of N=2 gauged supergravity and all relevant references in appendix \ref{conventions}.
In detail, we need three ingredients: the kinetic terms of the vectors in order to identify the gauge couplings, the gravitini-gauge vectors minimal couplings in order to identify the charge, and the value of the vacuum energy when the gravitino mass is vanishing. 
Since it is not restrictive, we assume that the gauging is purely electric. 
Once we establish that massless charged gravitini are in the swampland we show that if they have a parametrically small mass then the same results still apply. 

The kinetic terms of the gauge vectors $A_\mu^\Lambda$ have the form 
\begin{align}
e^{-1} {\cal L}_{kin.} = \frac14\, {\cal I}_{\Lambda \Sigma} \, F_{\mu\nu}^\Lambda F^{\mu\nu\,\Sigma} \, , 
\end{align}
where ${\cal I}_{\Lambda \Sigma}$ is a negative definite scalar dependent matrix and 
$F_{\mu\nu}^\Lambda = 2 \partial_{[\mu} A^\Lambda_{\nu]} + f^\Lambda{}_{\Sigma\Gamma} A^\Sigma_\mu A^\Gamma_\nu$. 
Once we define vielbeins and inverse vielbeins for the matrix ${\cal I}$ as follows 
\begin{align}
- {\cal I}_{\Lambda \Sigma} = \delta_{AB}\, {\cal E}^A_\Lambda {\cal E}^B_\Sigma \ , \qquad \quad {\cal E}^A_\Lambda {\cal E}^\Lambda_B = \delta_B^A \, , 
\end{align}
we get the kinetic terms for the canonical vectors $v^A = {\cal E}^A_\Lambda A^\Lambda$ 
\begin{align}
e^{-1} {\cal L}_{kin.} = - \frac14 \delta_{AB} F_{\mu\nu}^A F^{\mu\nu\,B} \, . 
\end{align} 
Within these $v^A_\mu$ vectors will be the massless U(1) gauge field we are interested in. 

Now we wish to identify the physical charge of the gravitini under this U(1). 
To this end we focus on the minimal coupling between the gravitini and the U(1) vector. 
The relevant term takes the form 
\begin{align}
\label{kin+coupl}
e^{-1} {\cal L}_{kin. \, 3/2} = 
- \overline \psi^i_\mu \gamma^{\mu\nu\rho} D_\nu(\omega) \psi_i{}_\rho  
- \frac{i}{2} \overline \psi^i_\mu \gamma^{\mu\nu\rho} v_\nu^A 
\left( \delta_i^j P_A^0 + \sigma^x{}_i{}^j P_A^x \right) 
\psi_j{}_\rho \, , 
\end{align}
once we define 
\begin{align}
P_\Lambda^0 {\cal E}^\Lambda_B = P^0_B \ , \qquad \quad P_\Lambda^x {\cal E}^\Lambda_B = P^x_B \,. 
\end{align} 
In \eqref{kin+coupl} we also include the kinetic term of the gravitino to stress that it is already canonically normalized. 
In choosing the vielbein basis we have enough freedom to ensure that our U(1) gauge field,  $u_m$, is along one specific basis element
\begin{align}
u_\mu  = v_\mu^{A=1}\,, 
\end{align}
so that the corresponding minimal coupling to the gravitino is
\begin{align}
e^{-1} {\cal L}_{kin. \, 3/2} = - \overline \psi^i_\mu \gamma^{\mu\nu\rho} D_\nu(\omega) \psi_i{}_\rho 
- i \overline \psi^i_\mu \gamma^{\mu\nu\rho} u_\nu Q_i{}^j \psi_j{}_\rho \, , 
\end{align}
where we defined the hermitian matrix 
\begin{align}
\label{grav-charge-Q}
2\, Q_i{}^j = \delta_i^j P_1^0 + \sigma^x{}_i{}^j P_1^x \, . 
\end{align} 
Since the two-by-two matrix $Q$ is hermitian we can diagonalize it by a unitary transformation $U$, which we can also use to rotate the gravitini
\begin{align} \label{su2rot}
Q \to U Q U^\dagger \ , \quad \psi \to U \psi \ , \quad \overline \psi \to \overline \psi U^\dagger \, , 
\end{align} 
so that the minimal coupling has the form
\begin{align}
e^{-1} {\cal L}_{kin. \, 3/2} = 
- \overline \psi^i_\mu \gamma^{\mu\nu\rho} D_\nu(\omega) \psi_i{}_\rho  
- i \overline{\psi}^1_\mu \gamma^{\mu\nu\rho} u_\nu q_1 \psi_1{}_\rho 
- i \overline{\psi}^2_\mu \gamma^{\mu\nu\rho} u_\nu q_2 \psi_2{}_\rho \, . 
\end{align}
Note that $q_1$ and $q_2$ are the physical couplings (i.e. gauge coupling $\times$ integer charge) between the canonical gauge bosons and the gravitini. 
Therefore they are the quantities that enter the WGC. 
As a result the magnetic WGC for the U(1) under which the gravitini are charged states that
\begin{align}
\Lambda_{UV} < q_1 \quad \& \quad \Lambda_{UV} < q_2 \,, 
\end{align}
where we remind the reader that we are working in Planck units. 

Now we turn to the scalar potential. 
We will see that under the assumption that the charged gravitini masses vanish, the vacuum energy hits the WGC cut-off. 
The N=2 scalar potential with vanishing gravitini masses takes the form 
\begin{align}
{\cal V} = - \frac12 {\cal I}^{-1|\Lambda \Sigma} \Big{[} P^0_\Lambda P^0_\Sigma + P^x_\Lambda P^x_\Sigma \Big{]} 
+ 4 h_{uv} \,k_\Lambda^u k_\Sigma^v\, \bar{L}^\Lambda L^\Sigma \, . 
\end{align}
This means the scalar potential has the property 
\begin{align}
{\cal V} \geq \frac12 \delta^{AB} \,\Big{[} P^0_A P^0_B + P^x_A P^x_B \Big{]} \, . 
\end{align}
Then we further have 
\begin{align}
\delta^{AB} \Big{[} P^0_A P^0_B + P^x_A P^x_B \Big{]} 
\!=\! \frac12 \delta^{AB} \Big{[} \delta_i^j P^0_A + \sigma^x{}_i{}^j P^x_A \Big{]} 
\Big{[} \delta_j^i P^0_B + \sigma^y{}_j{}^i P^y_B \Big{]} 
\!\geq\! \frac12 
\Big{[} \delta_i^j P^0_1 + \sigma^x{}_i{}^j P^x_1 \Big{]}^2 \,. 
\end{align}
Once we make use of the $Q$ matrix \eqref{grav-charge-Q} and perform the rotation \eqref{su2rot} we obtain
\begin{align}
{\cal V} \geq {\rm Tr} \left( U Q U^\dagger  U Q U^\dagger \right) = 
{\rm Tr} \left( Q Q \right) = q_1^2 + q_2^2 \,. 
\end{align}
We conclude that 
\begin{align}
\label{WGC-uplift}
{\cal V} \geq q_1^2 \quad \& \quad {\cal V} \geq q_2^2 \quad \Rightarrow \quad {\cal V} \geq \Lambda_{UV}^2 \,. 
\end{align}
This translates into 
\begin{align}
H \geq \Lambda_{UV} / \sqrt{3} \, , 
\end{align}
which means that the tree-level de Sitter critical points will receive large quantum corrections and cannot be trusted. 
This is a manifestation of the Dine--Seiberg problem \cite{Dine:1985he} and challenges the consistency of such de Sitter vacua. 
It is important to keep in mind that we are always talking about charges of the gravitino under massless gauge fields,  such that the weak gravity conjecture can be directly applied. 
Instead, when a gauge symmetry is broken even though a (covariantly) conserved current does still exist one cannot unambiguously define the charge any more, 
at least in Minkowski. 

As promised we can extend our conclusions to the case of very light gravitini, in particular when they are parametrically lighter than the Hubble scale. 
Indeed, a gravitino mass matrix has the form 
\begin{equation}
S_{ij} = i P_{\Lambda}^x L^\Lambda (\sigma_{x})_i^{\ k} \epsilon_{jk} \, , 
\end{equation}
and will only influence the supergravity scalar potential by the supersymmetry requirement that we include a new term of the form 
\begin{equation}
{\cal V}_{S} = - 4 \bar{L}^\Lambda L^\Sigma P^x_\Lambda P^x_\Sigma \,. 
\end{equation}
Having gravitino masses parametrically small compared to the Hubble scale means 
\begin{equation}
\sqrt{\bar{L}^\Lambda L^\Sigma P^x_\Lambda P^x_\Sigma} \ \ll \ H \,. 
\end{equation}
As a result the dominant contribution still comes from the term \eqref{WGC-uplift} and therefore the Hubble still hits the cut-off. 
We see that de Sitter backgrounds in N=2 supergravity with charged light gravitini are faced with a Dine--Seiberg problem. 
In the upcoming sections we will give explicit examples that show how the magnetic WGC restricts such vacua. 

Note that if the gravitini are uncharged the situation is different. 
Indeed in such a setup we would have $P_\Lambda^0=0=P_\Lambda^x$, and then the scalar potential takes the form 
\begin{align}
{\cal V} = 4 h_{uv} \,k_\Lambda^u k_\Sigma^v \,\bar{L}^\Lambda L^\Sigma \geq 0 \, . 
\end{align}
Thus, if we have an isometry with non-vanishing Killing vectors this can lead to positive vacuum energy while maintaining vanishing gravitini mass. 
However if such a background contains only spectator massless U(1) gauge fields then in any case the WGC cannot be directly applied  and we cannot conclude if it is in the swampland or not. 
We will present an example where this happens in subsection \ref{MUG}. 

If one does not consider gauged supergravities then the WGC is even less restrictive, at least at first sight. 
For example de Sitter vacua with an underlying non-linear realization of N=2 would not require charged gravitini or any gauging at all \cite{Kuzenko:2017zla,Antoniadis:2019hbu}. 
Such models can evade the restrictions we have found here but this does not mean they can arise from string theory, 
or even if they do, they may still lead to short-lived vacua \cite{Farakos:2020wfc}. 
In addition one may find complementary restrictions on such theories from EFT arguments as discussed in \cite{Jang:2020cbe}. 
There are also examples where the N=2 de Sitter is supported by condensates of gravitini bi-linears \cite{Kehagias:2009zz}, which hints that the vacuum does lie within a strongly coupled regime.

\subsection{Main result and related conjectures}

Our results have common ground with other conjectures and swampland bounds. 
It is thus instructive to state clearly what we find here and then discuss what is the relation to the existing swampland bounds. 
Our results here can be expressed in the following way: 
\begin{equation} 
\label{WGCdS}
\text{Quasi-de Sitter with} \quad m_{3/2} \ll H \quad \& \quad q_{3/2} \ne 0 \quad \text{has a Dine--Seiberg problem.} 
\end{equation} 
Indeed, when the conditions described in \eqref{WGCdS} are met we find that the EFT has a very low cut-off and so the two-derivative truncation is inherently inconsistent. 
One can thus say that such EFTs belong to the swampland. 
We have already presented a general proof for gauged N=2 in subsection \ref{chargedmassless}, and we also give a proof for N=8 in section \ref{sec:massless_gravitini_in_maximal_supergravity}. 
We further illustrate this result in the various examples in the following sections. 

Let us stress that the bound \eqref{WGCdS} follows from the magnetic weak gravity conjecture, and that it was already noticed in \cite{Cribiori:2020use} for gauged N=2 without hypermultiplets. 
There it was also rephrased as a conjecture, stating that de Sitter vacua with degenerate gravitino mass matrix belong to the swampland. 
Our results here thus yield further credence to such a bound, also in the presence of hypermultiplets. 

There is a non-trivial convergence between our results and the festina lente bound \cite{Montero:2019ekk}, which roughly states that $m^2 \gtrsim q g H$ has to hold for every charged particle in the spectrum, and has been further sharpened in \cite{Montero:2021otb}. 
There are three instances where we can draw compatible conclusions. 
Firstly, if we apply the FL bound on the gravitino we can bring it exactly to the form: 
\begin{align}
m_{3/2} \ll H \quad \& \quad q_{3/2} \ne 0 \quad \Longrightarrow \quad \text{in the swampland.}  
\end{align}
We see that this exactly matches our main conclusion. 
Thus our results can be considered solid independent evidence that the gravitino abides by the festina lente bound. 
Conversely, if we had assumed the FL bound, then \eqref{WGCdS} would emerge as simply a particular instance of it. 
Secondly, in \cite{Montero:2021otb} it is further argued that the Hubble is bounded from above by the magnetic WGC, which can be recast in a form that is relevant to us, that is: 
\begin{align}
H \gg q_{phys} M_P \quad \Longrightarrow \quad \text{in the swampland.}  
\end{align}
Again this condition is at the core of our work here and is already discussed in \cite{Cribiori:2020use}. 
Thirdly, according to \cite{Montero:2021otb} the FL bound also gives restrictions on non-abelian gauge theories, 
implying that they should either confine or break spontaneously at a scale above Hubble, that is: 
\begin{align}
\text{de Sitter with perturbative non-abelian gauging} \quad \Longrightarrow \quad \text{in the swampland.}  
\end{align}
This result again aligns nicely with our earlier discussion on non-abelian gaugings because we have used the WGC to argue that de Sitter N=2 vacua with perturbative non-abelian groups and massless gravitini are in the swampland. 

Clearly our work also makes partial contact with the de Sitter/TCC conjectures \cite{Danielsson:2018ztv,Obied:2018sgi,Andriot:2018wzk,Garg:2018reu,Bedroya:2019snp,Bedroya:2019tba,Ooguri:2018wrx} which indicate that de Sitter space either does not exist as a solution within a theory of quantum gravity, or is inherently unstable. 
Our results however differs in that it is based solely on the magnetic weak gravity conjecture without reference to the shape of the potential around the critical point. In particular, this leads to the elimination of certain de Sitter solutions that would be otherwise acceptable by the refined de Sitter conjecture \cite{Andriot:2018wzk,Garg:2018reu}, i.e.~de Sitter points with steep tachyons. 

Our analysis also makes contact with recent work \cite{Cribiori:2021gbf,Castellano:2021yye} claiming that the massless gravitini limit would correspond to a parametrically low cut-off due to towers of light states entering the EFT. 
Our work here and the earlier work \cite{Cribiori:2020use} is in agreement with these conjectures as the de Sitter points with vanishing gravitino mass are proposed to be in the swampland precisely because of a very low cut-off, and thus, clearly, so is the limit when approaching such points. 

\section{Stable de Sitter vacua with massless gravitini} \label{stable}

\subsection{SO(2,1) $\times$ U(1) with one hypermultiplet} 

The first illustrative model with vanishing gravitino masses comes from the gauging of a SO(2,1) $\times$ U(1) group in a supergravity model with three vector multiplets and one hypermultiplet.
The scalar manifolds are
\begin{equation}
		{\cal M}_{SK} = \left[\frac{{\rm{SU}(1,1)}}{\rm{U(1)}}\right]^3,  \qquad \qquad {\cal M}_{QK} = \frac{{\rm{SU}(2,1)}}{\rm{SU(2)}\times\rm{U(1)}}\,. 
\end{equation}
For the Special-K\"ahler geometry we use as a starting point the symplectic frame where the prepotential is 
\begin{equation}
		F(X) =  \sqrt{\left((X^0)^2 + (X^1)^2\right)\left((X^2)^2+(X^3)^2\right)} \, , 
\end{equation} 
which was shown in \cite{Fre:1991dm} to give a description in terms of the Calabi--Vesentini coordinates $z^I = \{S, y_0 , y_1\}$, 
by means of the symplectic sections 
\begin{align}
Z = \begin{pmatrix}  \frac{1}{2}( 1 + y_0^2 + y_1^2 ) \\[1mm] \frac{i}{2} (1 - y_0^2 - y_1^2) \\[1mm]  S y_0 \\[1mm] S y_1  \\[1mm] 
 \frac{1}{2}S ( 1 + y_0^2 + y_1^2 ) \\[1mm] \frac{i}{2} S (1 - y_0^2 - y_1^2) \\[1mm]  y_0 \\[1mm]  y_1   \end{pmatrix} \ .
\end{align}
The gauging that we perform is not electric in this frame and therefore we introduce the symplectic rotation
\begin{equation}\label{sp8}
		S_{\rm{Sp}(8.{\mathbb R})} = \left(\begin{array}{cccccc}
		 {\mathbb 1}_2&  &  &  &  & \\
		 & 0 & 0 &  & 1 & 0 \\
		 & 0 & -\sin \phi &  & 0 & \cos \phi\\
		 &  &  & {\mathbb 1}_2 &  & \\
		 & -1 &  0&  & 0 & 0 \\
		 & 0& -\cos \phi  &  & 0 &-\sin \phi
		\end{array}\right)
\end{equation} 
acting on the symplectic section according to (\ref{patches}).
The resulting holomorphic sections are
\begin{align}
Z = \begin{pmatrix}  \frac{1}{2}( 1 + y_0^2 + y_1^2 ) \\[1mm] \frac{i}{2} (1 - y_0^2 - y_1^2) \\[1mm] y_0 \\[1mm] y_1 (\cos~ \phi - S~ \sin~ \phi) \\[1mm] 
 \frac{1}{2}S ( 1 + y_0^2 + y_1^2 ) \\[1mm] \frac{i}{2} S (1 - y_0^2 - y_1^2) \\[1mm] - S y_0 \\[1mm] - y_1 ( S~ \cos~ \phi + \sin~ \phi) \end{pmatrix} \ ,
\end{align}
which fix the K\"ahler potential as
\begin{align}
e^{-\mathcal{K}} = -\text{Im} S \big( 1 - 2|y_0|^2 - 2|y_1|^2 + |y_0^2 + y_1^2|^2 \big) \ ,
\end{align}
and the rest of the geometry according to the formulae in the appendix \ref{conventions}.

In this parameterization there is an obvious SO(2,1) symmetry acting on the first three sections, generated by the Killing vectors
\begin{align}
\kappa_0^I = \begin{pmatrix} 0 \\[1mm] -\frac{i}{2} (1 + y_0^2 - y_1^2) \\[1mm] - i y_0 y_1 \end{pmatrix} \ , \quad
\kappa_1^I = \begin{pmatrix} 0 \\[1mm] \frac{1}{2}( 1 - y_0^2 - y_1^2) \\[1mm] - y_0 y_1 \end{pmatrix} \ , \quad 
\kappa_2^I = \begin{pmatrix} 0 \\[1mm] i y_0 \\[1mm] i y_1 \end{pmatrix},
\end{align}
which we choose to gauge with the first three vectors (the graviphoton and two of the other vectors in the vector multiplets), hence fixing
\begin{align}
k_\Lambda^I = e_0 \,\big(\kappa_0^I \ , \kappa_1^I \ ,  \kappa_2^I \ , \ 0 \big) \ ,
\end{align}
where we also introduced explicitly the SO(2,1) coupling $e_0$, which is going to be crucial in the following analysis.

The Quaternionic-K\"ahler geometry is that of the universal hypermultiplet (see for instance \cite{Ceresole:2001wi}), parametrized by the scalar fields $q^u = \{\rho, \sigma, \theta, \tau\}$, with metric
\begin{align}
ds^2 = h_{uv} dq^u dq^v = \frac{d\rho^2}{2 \rho^2} + \frac{1}{2 \rho^2} (d\sigma - 2 \tau d\theta + 2 \theta d \tau)^2 + \frac{2}{\rho} (d\theta^2 + d \tau^2) \ .
\end{align}
In this sector we decide to gauge a compact U(1) symmetry generated by the Killing vector
\begin{align}
\kappa_H^u = \begin{pmatrix} 4 \rho \tau \\[1mm] 2 \theta + 2 \sigma \tau + 2 \rho \theta + 2 \theta(\theta^2 + \tau^2) \\[1mm] 4 \theta \tau - \sigma \\[1mm]  1 - \rho - 3 \theta^2 + \tau^2 \end{pmatrix} \ .
\end{align}
The gauging is performed using the last vector available, hence fixing
\begin{align}
k_\Lambda^u = e_1\, \big( 0 \ , \ 0 \ , \ 0 \ ,  \kappa_H^u \big) \ ,
\end{align}
where once again, we made explicit the coupling $e_1$.
Since the scalar potential (\ref{pot}) is determined not only by the Killing vectors of the hypermultiplets, but also by their prepotentials, we also give here the explicit form of the prepotential associated to the isometry $\kappa_H^u$:
\begin{align}
P_3^x = e_1 \begin{pmatrix} -\frac{2}{\sqrt{\rho}}(1+\rho-3\theta^2 + \tau^2) \\[1mm]  \frac{2}{\rho} (\sigma - 4 \theta \tau) \\[1mm]  \frac{2}{\rho} ( \theta + 3 \rho \theta - \theta^3 + \sigma \tau - \theta \tau^2)  \end{pmatrix} \ .
\end{align}

Once one puts together the various pieces to the scalar potential, one can see that it has a critical point at
\begin{align}
S = \cot \phi - \frac{i}{4} \left|\frac{e_0}{e_1 \sin \phi} \right|\ , \quad \rho = 1 \ , \quad y_0 = y_1 = \sigma = \theta = \tau = 0 \ ,
\end{align}
where 
\begin{align}
{\cal V} = 4~| e_0~ e_1~ \sin \phi|. 
\end{align}
This implies that the Hubble scale at this critical point is 
\begin{equation}	
		H = \sqrt{\frac{4}{3} |e_0 e_1 \sin \phi|} \ .
\end{equation}
Moreover the gravitini mass matrix is identically vanishing at this critical point.
The U(1) Killing vector also vanishes at this critical point, indicating that the U(1) symmetry is preserved and thus the WGC can be applied.

To check explicitly the consistency of such vacua against the weak gravity conjecture, we first compute the gauge couplings at the critical point, which follow from 
\begin{align}
\mathcal{I}^{-1|\Lambda \Sigma} =  - \frac{1}{2} \sin \phi \begin{pmatrix} 4 e_1/e_0 & 0 & 0 & 0 \\[1mm] 
																						0 & 4 e_1/e_0 & 0 & 0 \\[1mm] 
																						0 & 0 & 4 e_1 / e_0 & 0 \\[1mm] 
																						0 & 0 & 0 & e_0 / e_1 \end{pmatrix} \ .
\end{align}
We also note that the gravitino is only charged under the U(1) symmetry with charge $q_{3/2} = 2 e_1$, so the magnetic WGC cut-off is
\begin{align}
\Lambda_{UV} = g_{U(1)} q_{3/2} = \sqrt{2 |e_0 e_1 \sin \phi |} \ ,
\end{align}
and we see that the Hubble scale is of the order of the cut-off dictated by the magnetic WGC and illustrates why there is a Dine--Seiberg problem. 

The mass spectrum of the scalar fluctuations around the critical point includes two zero-modes, corresponding to the goldstone modes eaten by the two broken non-compact SO(2,1) isometries. 
The rest of the spectrum is positive definite 
\begin{align}
m^2_{(multiplicity)} = \big( 0_{(2)} \ , \ 1/4_{(4)} \ , \ 1_{(2)} \ , \ 2_{(2)} \big) \times {\cal V} \, , 
\end{align}
so this critical point is also in violation of the de Sitter criterion.

For completeness, we note that there is also another unbroken U(1) isometry coming from the compact generator of $SO(2,1)$ gauged on the vectors. 
The gravitino charge under that U(1) is given by $P^0_2$, which at the critical point simply evaluates to
\begin{align}
q_{3/2} = \frac{1}{2} P^0_2 = e_0 \,. 
\end{align}
Multiplying by the appropriate component of $\mathcal{I}$ we obtain the same cut-off as before, $\Lambda_{UV} =  \sqrt{2 |e_0 e_1 \sin \ \phi|}$, 
which again points to the Dine--Seiberg problem.

\subsection{SO(2,1) $\times$ U(1)$^3$ with two hypermultiplets} 

The model presented in this subsection has again massless gravitini at its critical point. 
A version of this model without hypermultiplets can be found in \cite{Fre:2002pd} and is already discussed from the WGC perspective in \cite{Cribiori:2020use} and eliminated. 

The modification that we consider here contains two hypermultiplets and is the first time a model with two hypermultiplets and a fully stable de Sitter vacuum has been constructed. However, as we will see, it still suffers from a Dine--Seiberg problem that is signaled by the WGC. 

The matter content of the model is given by five vector multiplets and two hypermultiplets, with scalar geometry
\begin{equation}\label{scalarSO42}
		{\cal M}_{SK} = \frac{{\rm{SU}(1,1)}}{\rm{U(1)}} \times \frac{{\rm{SO}(2,4)}}{\rm{SO(2)}\times\rm{SO(4)}},  \qquad \qquad {\cal M}_{QK} = \frac{{\rm{SO}(4,2)}}{\rm{SO(4)}\times \rm{SO(2)}}\,.
\end{equation}

\subsubsection{Vector multiplets}

The geometry of the vector multiplet sector is described in a similar way as in the example above, starting from the prepotential
\begin{equation}
		F(X) =  \sqrt{\left((X^0)^2 + (X^1)^2\right)\left(X^{\tilde a} X^{\tilde b} \delta_{{\tilde a}{\tilde b}}\right)},
\end{equation}
where ${\tilde a},{\tilde b}=2,3,4,5$.
As in the previous case, we can describe our gauging in the electric frame by introducing the Calabi--Vesentini coordinates \cite{Fre:1991dm,Fre:2002pd}
\begin{align}
z^I = \{S \, , \ y^a \}\,,  \qquad {\rm{where}}\qquad y^a = \{y^0, y^x\}, \quad x=1,2,3 \, , 
\end{align}
and by performing an appropriate symplectic rotation analogous to (\ref{sp8}). 

The resulting holomorphic sections in the new frame are
\begin{equation} \label{hol-sec}
Z = 
\begin{pmatrix} X^\Lambda \\[1mm] F_\Sigma \end{pmatrix} \, , 
\end{equation}
where 
\begin{align}
X^\Lambda(S,y^a) = \begin{pmatrix} 
\frac12 (1 + y^a y^a  ) 
\\[1mm] 
\frac{i}2 (1 - y^a y^a  ) 
\\[1mm] 
y^0
\\[1mm]
y^x (\cos \phi - S \sin \phi)
\end{pmatrix} \, , 
\end{align}
and
\begin{align}
F_\Lambda(S,y^a) = \begin{pmatrix} 
\frac12 S (1 + y^a y^a  ) 
\\[1mm] 
\frac{i}2 S (1 -y^a y^a  ) 
\\[1mm] 
-S y^0
\\[1mm]
-y^x( S \cos \phi + \sin \phi)
\end{pmatrix} \,. 
\end{align}
The geometry of this sector follows from these sections according to the formulae in appendix \ref{conventions}. 
The K\"ahler potential is the sum of two factors
\begin{align}
{\cal K} = - \log [ i (\overline X^\Lambda F_\Lambda - \overline F_\Sigma X^\Sigma) ] 
= {\cal K}_1 + {\cal K}_2 \, , 
\end{align}
where 
\begin{align}
{\cal K}_1 &= - \log[ i (S - \overline S) ] \, , 
\\
{\cal K}_2 &= - \log\left[ \frac12 ( 1 - 2 y^a \overline y^a + y^a y^a \overline y^a \overline y^a ) \right] \,. 
\end{align}
The metric for the scalar fields is factorized as 
\begin{align}
g_{I \overline{J}} = \begin{pmatrix} g_{S \overline S} & 0 \\ 0 & g_{a \overline b} \end{pmatrix} \,, 
\end{align}
with
\begin{align}
g_{S \overline S} = \frac{1}{(2 \, \text{Im} S)^2} \ , \quad 
g_{a \overline b} = \frac{\partial}{\partial y^a} \frac{\partial}{\partial \overline y^b} {\cal K}_2 \,. 
\end{align}

Also in this model we gauge the SO(2,1) symmetry in the vector sector that rotates the first three sections.
This is generated by the Killing vectors
\begin{align}
\kappa_0^I & = \left(0,-\frac{i}{2} \left[1 + y_0^2 - \sum_x (y_x)^2 \right] , - i y_0 y_1 , - i y_0 y_2 , - i y_0 y_3 \right) \,, 
\\[2mm]
\kappa_1^I & = \left(0,\frac{1}{2} \left[1 - y_0^2 + \sum_x (y_x)^2 \right], - y_0 y_1 , - y_0 y_2 , - y_0 y_3 \right) \,, 
\\[2mm]
\kappa_2^I & = \left(0, i y_0 , i y_1 , i y_2 , i y_3 \right) \,.
\end{align}
We gauge these isometries with the graviphoton and the first two vectors in the vector multiplet sector. 
The Killing vectors then are
\begin{align}\label{killing1}
k_\Lambda^I = e_0 \,\big( \kappa_0^I \ , \kappa_1^I \ ,\kappa_2^I \ , \ 0 \ , \ 0 \ ,\ 0 \big),
\end{align}
where we made once more explicit the coupling $e_0$.
From the above ingredients we can compute, 
\begin{align}
{\cal V}_{D_1} = g_{I \overline{J}} \, k_\Lambda^I k_\Sigma^{\overline{J}} \, \overline{L}^\Lambda L^\Sigma \, , 
\end{align}
which only depends on the vector multiplets and will therefore remain the same regardless of our choice of hypermultiplets or their gauging.

\subsubsection{Gauging U(1)$^3$ on two hypermultiplets} 

We will now include two hypermultiplets in the model that we studied in the previous subsection. 
The hyper manifold ${\cal M}_{QK}$ given above is a coset space and therefore we can exploit this fact to explicitly provide the details of its construction in the appendix \ref{so42}. 
We only report here its metric
\begin{eqnarray}
\nonumber 
		ds^2 \!\!&=&\!\!\! h_{uv} dq^u dq^v 
		\\[1mm]
		\nonumber 
		&=&\!\!\! \frac{1}{q_1^2} \left[ dq_1^2 + q_5^2 dq_4^2 + (dq_2 + \sqrt{2} q_7 dq_4)^2 + (dq_3 + \sqrt{2} q_8 dq_4)^2\right]  
		\\[1mm]
		\nonumber 
		&&\!\!\! + \frac{1}{ 72 q_1^2 q_5^2} \left[6 \sqrt{2} dq_6 - 12 q_7 dq_2 - 12 q_8 dq_3 + 2 \sqrt{2} q_4 (q_7 dq_7 + q_8 dq_8) - 5 \sqrt{2} (q_7^2 + q_8^2) dq_4) \right]^2 
		 \\[1mm]
		&&\!\!\! + \frac{1}{q_5^2} \big(dq_5^2 + dq_7^2 + dq_8^2 \big) 
		\,.  
\end{eqnarray}

Given that the isometries of ${\cal M}_{QK}$ are a subset of those of ${\cal M}_{SK}$, one could gauge an $SO(2,1) \times SO(3)$ gauge group using at the same time their action on the vector scalars and on the hypers.
This is what was done in \cite{Fre:2002pd} to find one of the first examples of marginally stable de Sitter vacua. 
These models however do not lead to scalars with all masses positive 
and a simple analysis of their vacuum structure also shows that they are in tension with the WGC. 

We therefore decided to follow a different path and gauge three abelian commuting isometries in the hypermultiplet geometry while leaving the SO(2,1) action confined to the vector multiplet sectors.
To summarize, our gauging is 
\begin{align}
{\cal G}_{gauge} = SO(2,1)_{only \ on \ vectors} \times \big{(} U(1) \times U(1) \times U(1) \big{)}_{only \ on \ hypers} \,.
\end{align}
This gauging is specified by the Killing vectors (\ref{killing1}), together with the Killing vectors specifying the isometries that we want to gauge on the hypermultiplet sector 
\begin{align}
U(1)^3 : \,k_\Lambda^u = \big(0 \ , \ 0 \ , \ 0 \ , \ e_4\, k_{T_{\underline{12}}}^u \ , \ e_5\, k_{T_{\underline{34}}}^u \ , \ e_6\, k_{T_{\underline{56}}}^u \big) \,,
\end{align}
where the explicit expression of the $k_{T_{\underline{ab}}}^u$ can be found in appendix \ref{so42}.
The gauging is electric in the frame given by \eqref{hol-sec}. 
Notice that we could have made use of additional symplectic rotation parameters $\phi_i$ for each U(1) and indeed that would lead to different expressions for the masses. 
However, in all cases where the masses are positive, the properties of the vacuum do not significantly depend on the values of the angles and so we took them here to be all of the same value $\phi$ for simplicity. 
In the end these ingredients contribute to the scalar potential that relates to the hypers and has the form 
\begin{align}
{\cal V}_{D_2} = 4\, h_{uv} \,k_\Lambda^u k^v_\Sigma \overline L^\Lambda L^\Sigma \,. 
\end{align}

When hypers are introduced the would-be FI terms are field-dependent and are given by the appropriate prepotentials $P_\Lambda^x (q^u) $, which are determined by the isometries gauged on the hypers as reviewed in appendix \ref{conventions}. 
We thus have 
\begin{align}
{\cal V}_{F} = \left(  U^{\Lambda \Sigma} 
- 3 \overline{L}^\Lambda  L^\Sigma \right) P_\Lambda^x P_\Sigma^x \,, 
\end{align} 
and the total potential is ${\cal V}_{D_1}+{\cal V}_{D_2}+{\cal V}_{F}$. 

One can then verify that there is a central critical point at
\begin{align}
q^1=q^5=1 \ , \quad  q^2=q^3=q^4=q^6=q^7=q^8=0 \ , 
\end{align}
and 
\begin{align}
y^0=y^x=0 \ , \quad S = \cot \phi - i \left|\frac{e_0}{\sqrt{e_4^2 + e_5^2 } \, \sin \phi}\right| \, . 
\end{align}
It is interesting to note that at this critical point most prepotentials vanish
\begin{align}
P_0^x = P_1^x = P_2^x = P_5^x= \begin{pmatrix} 0 \\ 0 \\ 0 \end{pmatrix} \ , \quad 
P_3^x = \begin{pmatrix} -e_4 \\ 0 \\ 0 \end{pmatrix} \ , \quad 
P_4^x = \begin{pmatrix} -e_5 \\ 0 \\ 0 \end{pmatrix} \,, 
\end{align}
as well as all Killing vectors of the compact U(1) isometries gauged in the hyper-sector 
\begin{align}
k_\Lambda^u=0 \,. 
\end{align}
This is in accordance with the fact that we have a residual U(1)$^4$ gauge symmetry on the vacuum.
The value of the scalar potential is then 
\begin{align}\label{V000}
{\cal V} =  \sqrt{e_4^2 + e_5^2 } \, |e_0 \sin\phi| \, , 
\end{align}
and the canonically normalized mass eigenvalues are given by 
\begin{align}
m^2_{(multiplicity)} = \left( 0_{(2)}\ , \ 1_{(6)}\ , \ 2_{(2)}\ ,\ \frac{e_4^2}{e_4^2 + e_5^2}{}_{(4)} \ ,\ \frac{e_5^2}{e_4^2 + e_5^2}{}_{(4)} \right) \times {\cal V} \,,
\end{align} 
which include two goldstone modes, while all the other masses are positive definite.
We therefore see that we have a fully stabilized de Sitter critical point with both vector and hypermultiplets. 
This is the first instance where a model with these properties is constructed. 
We also see however that the gravitini remain massless 
\begin{align}
S_{ij} = i P_{\Lambda}^x L^\Lambda (\sigma_{x})_i^{\ k} \epsilon_{jk} = 0 
\end{align}
and hence we expect this model to fail to give a proper effective theory, according to the WGC.

In order to check this, we need the kinetic terms of the vectors, 
and in particular of the ones that are performing the U(1) gaugings. 
At the critical point this sector of the Lagrangian becomes
\begin{align}
e^{-1} {\cal L}_{kin. vec.} = - \frac14 \, \left|\frac{e_0}{\sqrt{e_4^2 + e_5^2 } \, \sin\phi}\right| \, \sum_{\Lambda=0}^2 F^2(A^\Lambda) 
- \frac14 \, \frac{\sqrt{e_4^2 + e_5^2 }}{|e_0 \sin\phi|} \, \sum_{\Lambda=3}^5 F^2(A^\Lambda) \,
\end{align}
and we see that there is a rather intricated dependence of the gauge couplings on the charges $e_0$, $e_4$ and $e_5$.
The simplest way to check compatibility with the magnetic WGC is the following. 
We first notice that we have a spontaneous breaking of the SO(2,1) to a U(1) and the goldstone modes associated to this symmetry breaking are the real and imaginary parts of $y_0$. 
This is seen from the fact that on the vacuum we have 
\begin{align}
k_0^I & = e_0 \left(0,-\frac{i}{2} , 0 , 0 , 0 \right) \,, 
\\[2mm]
k_1^I & = e_0 \left(0,\frac{1}{2} , 0 , 0 , 0 \right) \,, 
\\[2mm]
k_2^I & = e_0 \left(0, 0 , 0 , 0 , 0 \right) \, , \\[2mm]
k_3^I &= k_4^I = k_5^I = 0 \,. 
\end{align} 
This means that the U(1) that survives the Higgsing is the one corresponding to $k_2^I$, which is just a standard U(1) acting on the $y^x$'s as follows 
\begin{align}
U(1)_{residual} \ : \ y^x \rightarrow \, e^{i \alpha \, e_0} \, y^x \,. 
\end{align}
As a result we can identify the physical minimal coupling between the $y^x$'s and the residual massless U(1) gauge vector as 
\begin{align}\label{qphys0}
q_{phys} =e_0 \times \sqrt{ \sqrt{e_4^2 + e_5^2 } \,\left|\frac{ \sin\phi}{e_0} \right| } \,. 
\end{align} 
For any charged field the magnetic WGC tells us that in Planck units 
\begin{align}
\Lambda_{UV} < q_{phys} \,,
\end{align}
while inspecting (\ref{V000}) and (\ref{qphys0}) we have that
\begin{align} 
H \sim \Lambda_{UV} \,, 
\end{align}
which is the signal that such vacua are faced with a Dine--Seiberg problem. 
One can reach the same conclusion by identifying the gravitino charge under the residual U(1) from $P_2^0/2$ which gives $q_{3/2} = e_0$, 
and thus also the gravitino has physical coupling \eqref{qphys0}. 

It is interesting to note that the model we just presented can also be obtained from a reduction from the SO(4,4) gauged N=8 supergravity of \cite{DallAgata:2012plb}.
In particular, if we set $e_5=e_6=0$ in the model above and keep only the $e_4$ we get that the mass eigenvalues are given by 
\begin{align}
m^2_{(multiplicity \, + \, goldstones)} = \left( 0_{(4+2)} , 1_{(10)},2_{(2)} \right) \times {\cal V} \,, 
\end{align} 
which match the truncated spectrum of the central vacuum in \cite{DallAgata:2012plb}. 
This can be understood from the fact that the scalar manifold (\ref{scalarSO42}) can be obtained as a N=2 truncation of the N=8 scalar manifold E$_{7(7)}/$SU(8), following \cite{Andrianopoli:2002rm}, and that the SO(4,4) gauging produces an action on the scalar fields which is factorized in the same way as that of our truncated model.

\subsubsection{Gauging the SO(3)} 

The same model with scalar manifold (\ref{scalarSO42}), has been used in \cite{Fre:2002pd}, but with a SO(3) gauging rather than a U(1)$^3$.
Also, both the SO(3) and the SO(2,1) factors have been gauged with a diagonal action on the vectors and hypermultiplets.
The resulting scalar potential has a critical point where the hypers have non-negative masses.
For completeness, we show here that these models are still faced with a Dine--Seiberg problem. 

Since all details of the model can be found in \cite{Fre:2002pd}, we will only report here the details relevant for our discussion.
Let us recall that $e_0$ corresponds to the coupling of the SO(2,1) factor (and $r_0=0,1$ is the coefficient that signals the presence of a simultaneous action on the hyperscalars) and $e_1$ is the coupling of the SO(3) factor (and $r_1 = 0,1$ again signals the action on the hyperscalars).

These models have a meta-stable vacuum with vacuum energy
\begin{align}
{\cal V} = \sqrt{ 3 (1 + 2 r_0^2) } \, |e_0 \, e_1 \, r_1 \, \sin\phi| > 0 \,. 
\end{align}
On this point the SO(2,1) gauge group is broken to a residual U(1), whose gauge vector has a kinetic term of the form
\begin{align}
e^{-1} {\cal L}_{residual \, U(1)} = - \frac14 \left| \frac{e_0 \, \sqrt{(1 + 2 r_0^2) }}{\sqrt{ 3 } \, e_1 \, r_1 \, \sin\phi} \right| F_{\mu\nu} F^{\mu\nu} \,. 
\end{align}
Under the surviving U(1) the scalars of the vector multiplet are still charged with charge $e_0$, 
that is 
\begin{align}
U(1)_{residual} \ : \delta y^x = i \alpha \, e_0 \, y^x \,. 
\end{align}
As a result the physical charge of the $y^x$ scalars under the residual U(1) is 
\begin{align}
q_{phys} =\sqrt{ \left|\frac{\sqrt{ 3 } \, e_1 \, r_1 \, \sin\phi}{e_0 \, \sqrt{(1 + 2 r_0^2) }}\right|} \times e_0 \, , 
\end{align}
which sets the upper bound on the WGC cut-off. 
We conclude again that 
\begin{align}
H \sim \Lambda_{UV} \,, 
\end{align}
both for $r_0=0$ and for $r_0=1$. 
Note that here we used the charge of the $y^x$ scalars under the U(1) to find the WGC cut-off, 
but we could have equally well used the gravitino charge. 

Note that in all of the examples presented in this section, the central vacuum has vanishing gravitino mass. The contribution to the gravitino mass from the SO(2,1) gauging vanishes because the corresponding prepotentials vanish, while the contributions associated to the rest of the gauge group vanish due to the vanishing of the corresponding section components. Thus all these examples serve to illustrate the result that critical points with charged massless gravitini violate the WGC. Of course, these examples also violate the de Sitter criterion directly, by virtue of their scalar mass spectra being positive (semi-)definite.

\section{Multiple unstable de Sitter vacua with various gravitini masses} 
\label{so3}

We now turn to a different set of examples, where the de Sitter critical points are unstable and could survive the de Sitter conjecture. 
We will show that whenever the gravitini masses are vanishing, we still can place these models in the swampland.
Also, among these examples we find a case where there is a modulus such that the gravitini masses vary with its expectation value.
This is a very instructive example because it shows explicitly how we can violate the WGC in a dynamic way, 
showing that not only vanishing gravitini masses are dangerous, but also very light ones. 
We also give another example that we think is instructive, because it makes explicit the discussion of our section \ref{non-abelian}, having a central vacuum with non-abelian gauge symmetry and massless gravitini, for which the argument in section \ref{non-abelian} allows to point out a Dine--Seiberg problem.

\subsection{Scalar manifolds}

Both models we consider in the following contain three vector multiplets and two hypermultiplets, parameterizing the scalar manifold
\begin{equation}\label{scalar2man}
		{\cal M}_{SK} = \frac{{\rm{SU}(1,3)}}{\rm{SU(3)}\times\rm{U(1)}},  \qquad \qquad {\cal M}_{QK} = \frac{{\rm{SO}(4,2)}}{\rm{SO(4)}\times \rm{SO(2)}}\,.
\end{equation}
The Special-K\"ahler manifold describes a vector multiplet geometry with minimal couplings and follows from the prepotential
\begin{equation}
		F(X) = -\frac{i}{4} X^\Lambda X^\Sigma \, \eta_{\Lambda \Sigma},
\end{equation}
with $\eta = \rm{diag}\{1,-1,-1,-1\}$.
The associated symplectic frame is described by the holomorphic sections
\begin{align}
Z = \begin{pmatrix} 1 \\[1mm] z^I \\[1mm] -\frac{i}{2} \\[1mm] \frac{i}{2} z^I \end{pmatrix}, \qquad I = 1,2,3 \, , 
\end{align}
where the $z^I$ are the three complex vector multiplet scalars. 
The K\"ahler potential is given by
\begin{align}
\mathcal{K} = - \log\left[ 1 - z^I \bar{z}^I \right],
\end{align}
which makes explicit the $SU(2)$ isometry that rotates the three scalars. 
The Killing vectors for these isometries are
\begin{align} \label{veckills}
\kappa_1^I = \begin{pmatrix} 0 \\ z_3 \\ - z_2  \end{pmatrix} \ , \quad  \kappa_2^I = \begin{pmatrix} -z_3 \\ 0 \\ z_1  \end{pmatrix} \ , \quad  \kappa_3^I = \begin{pmatrix} z_2 \\ -z_1 \\ 0  \end{pmatrix} \,. 
\end{align}

The hypermultiplet scalar manifold is the same as in the previous section and the details of its parametrization are given in appendix \ref{so42}.

\subsection{Gauging}

The two models that we are going to analyze in the following have a gauge group that is the direct product of an SO(3) factor and an abelian compact or non-compact one-dimensional group.

The common SO(3) factor is taken to act simultaneously on the vector multiplet scalars as well as on the hyperscalars.
The action on the vector scalar fields is identified with the isometries generated by the Killing vectors \eqref{veckills} and is gauged by the vector fields in the vector multiplets 
\begin{align}
k_\Lambda^I = e_1\, \big( 0 \ , \ \kappa_1^I \ , \ \kappa_2^I \ , \ \kappa_3^I \big) \,. 
\end{align}
The same SO(3) gauge group acts on the hyperscalars as specified by the generators $T_{\underline{12}}, T_{\underline{13}} , T_{\underline{23}}$ of the $\mathfrak{so}(4,2)$ algebra (see equation \eqref{so42gens} in appendix \ref{so42}). 
In addition we either take a compact abelian factor gauged by the graviphoton and acting on the hyperscalars as specified by $T_{\underline{56}}$, or a non-compact abelian factor, always gauged by the graviphoton, and acting on the hyperscalars as specified by $T_{\underline{46}}$.
Overall on the hypermultiplets we have the identifications
\begin{align}
SO(3) \times U(1) : \,k_\Lambda^u = \big( e_0 k_{T_{\underline{56}}}^u \ , \ e_1 k_{T_{\underline{12}}}^u \ , \ e_1 k_{T_{\underline{13}}}^u \ , \ e_1 k_{T_{\underline{23}}}^u \big), \\[2mm]
SO(3) \times O(1,1) : \,k_\Lambda^u = \big( e_0 k_{T_{\underline{46}}}^u \ , \ e_1 k_{T_{\underline{12}}}^u \ , \ e_1 k_{T_{\underline{13}}}^u \ , \ e_1 k_{T_{\underline{23}}}^u \big),
\end{align}
i.e. the SU(2) acting on the vector multiplets is identified with the SO(3) of the hyper manifold, while the U(1) or the O(1,1) symmetry is gauged by the graviphoton. 
{}From the Killing vectors and the metric we can also compute the prepotentials $P^x$ using \eqref{Psoln}.

Both models have a critical point at the SO(3) invariant point
\begin{align}\label{critcpoint}
q_1 = q_5 = 1 \ , \quad  q_2 = q_3 = q_4 = q_6 = q_7 = q_8 = 0 \ , \quad z^I = 0,
\end{align}
where the SO(3) Killing vectors vanish. The U(1) Killing vector also vanishes at this point, hence showing that the whole gauge group survives, while the O(1,1) Killing vector takes the form 
\begin{align}
k_{\underline{46}}^u = \delta^u_8,
\end{align}
thus signalling its breaking at the critical point.
The corresponding prepotentials at the same point are 
\begin{align}
P_0^x = \begin{pmatrix} 0 \\ 0 \\ 0 \end{pmatrix} \ , \quad P_1^x = \begin{pmatrix} e_1 \\ 0 \\ 0 \end{pmatrix} \ , \quad P_2^x = \begin{pmatrix} 0 \\ e_1 \\ 0 \end{pmatrix} \ , \quad P_3^x = \begin{pmatrix} 0 \\ 0 \\ e_1 \end{pmatrix} \ ,
\end{align}
with $P_0^x$ vanishing for both the U(1) and the O(1,1) generators. 
As anticipated in the introduction to this section, both models have interesting features for our analysis, which we now are going to examine.

\subsection{SO(3) $\times$ U(1)}

The central critical point in this model has energy 
\begin{align}
{\cal V} = 3 \,e_1^2,
\end{align}
while the eigenvalues of the scalar mass matrix are
\begin{align}
m^2_{(multiplicity)}=\left(-\frac{2}{3}_{(6)} \ , \ \frac{4}{3}r^2_{(2)} \ , \ \frac{4}{3}(r^2+1)_{(6)}\right) \times {\cal V},
\end{align}
where $r = e_0/e_1$ is the ratio of the U(1) and SO(3) couplings. 
The gravitino mass matrix vanishes at this critical point and hence we could be within the assumptions of our general proof of section \ref{chargedmassless}.
However, the gravitino charges under the four gauge bosons are 
\begin{align}
q_{A} = \left( \pm 0 \ , \ \pm \frac{1}{\sqrt{2}} e_1 \ , \ \pm \frac{1}{\sqrt{2}} e_1 \ , \ \pm \frac{1}{\sqrt{2}}e_1 \right),
\end{align}
so that we see that the gravitini are not charged under the U(1), but only with respect to the SO(3) gauge group.
The charges listed above have been computed by taking into account the normalization of the vector kinetic terms, given by  the values of the gauge kinetic functions at the critical point, namely ${\cal I}_{\Lambda \Sigma} = -\frac12 \, \delta_{\Lambda \Sigma}$.

Since the gravitini are not charged under the U(1) gauge group, and the SO(3) factor is unbroken, we can not confidently apply the WGC in its usual form, using its gauge coupling. 
We must therefore resort to the argument presented in section \ref{non-abelian} where we look at nearby configurations that break the SO(3) symmetry down to a U(1). 
This allows us to use the SO(3) coupling in the magnetic WGC, giving a cut-off $\Lambda_{UV} = \frac{e_1}{\sqrt{2}}$, and the Hubble scale exceeds this. 
In particular, we assume a small perturbation of the SO(3) point of the form 
\begin{align}
\label{un-critical}
z_2 = i \, \epsilon \ , \quad z_1=z_3=0 \,. 
\end{align}
This point is not a critical point of the theory of course, but it is still a legitimate configuration in our field space. 
In particular since Im$\,z_2$ gets a vacuum expectation value, the central SO(3) gauge group is Higgsed and only a U(1) remains under which fields are still charged with charge $e_1$. 
In addition the total energy density, which is dominated by the vacuum energy, is still given approximately by $\rho \simeq 3\, e_1^2 + \epsilon \,\frac{d V}{d \text{Im} z_2}$. 
As a result, for small $\epsilon$ we have 
\begin{align}
\label{dev-WGC}
\Lambda_{WGC}\Big{|}_{\epsilon \sim 0} \sim e_1 \sim H \,, 
\end{align}
which can be extrapolated to the central vacuum as the limit $\epsilon \to 0$. 
We conclude that the central critical point is also threatened by the WGC cut-off.

It is worth pointing out that for non-zero $\text{Im}\, z_2$, the derivative of the potential also points in that direction.
This allows for classical trajectories which preserve the U(1) symmetry throughout their entire duration. 
This puts us in the second scenario discussed in section \ref{non-abelian}, where we do not have a modulus, 
but do still have a classical path. 
An example, where $\text{Im}\, z_2$ is a true modulus will be however presented in the next subsection.

Before moving on, we would like to point out that the same model possesses a second critical point, where the SO(3) gauge group is fully broken.
This vacuum can be found by letting, for example, $\text{Re}\, z_1$ and $\text{Im}\, z_2$ vary. 
The new critical point appears at
\begin{align}
\text{Re}\, z_1 = \frac{1}{2} \ , \quad \text{Im}\, z_2 = \frac{1}{2},
\end{align}
and has energy ${\cal V} = 2\, e_1^2$. 
The normalized scalar mass spectrum is
\begin{align}
m^2_{(mult.)}&=\left(0_{(3)}, -1_{(2)} , 8_{(1)},2+4r^2-2r_{(2)},\beta^2+\beta_{(2)}, \beta^2 - \beta_{(2)} , 2+2r+4r^2_{(2)}\right) \times {\cal V} \, , 
\end{align}
with $r=e_0/e_1$ describing the ratio of the charges and $\beta = 4 r + 1$.
Once again this critical point respects the de Sitter criterion and also our criterion fails because the gravitino mass matrix at this saddle point is
\begin{align}
S_{ij} = \begin{pmatrix} \sqrt{2}\, e_1 & 0 \\ 0 & 0 \end{pmatrix} \, , 
\end{align}
so that one gravitino acquires a mass of order the Hubble scale, while the other remains massless.
We can also see that the gravitini are still uncharged under the residual U(1), because the gauge kinetic functions at this point are given by
\begin{align}
-\frac{1}{2}\mathcal{I}^{-1} = \begin{pmatrix} 3 & 2 & 0 & 0 \\
											  2 & 2 & 0 & 0 \\
												0 & 0 & 2 & 0 \\
												0 & 0 & 0 & 1 \end{pmatrix}
\end{align}
and physical gravitino charges are then given by the eigenvalues of
\begin{align}
(q_A)_i{}^j = \frac{1}{2}\mathcal{E}_A^\Lambda P_\Lambda^x (\sigma^x)_i{}^j,
\end{align}
where 
\begin{align}
\mathcal{E}_{\ A}^\Lambda = \begin{pmatrix}
\sqrt{2} & 2 & 0 & 0 \\
0 & 2 & 0 & 0 \\
0 & 0 & 2 & 0 \\
0 & 0 & 0 & \sqrt{2} 
\end{pmatrix},
\end{align} 
so that $\mathcal{I}^{-1|\Lambda \Sigma} = \mathcal{E}_{ \ A}^\Lambda \mathcal{E}_{ \ B}^\Sigma \delta^{AB}$.
After a straightforward calculation we find that 
\begin{align}
q_{A} = \left( \pm 0 \ , \ \pm e_1 \ , \ \pm e_1 \ , \ \pm \frac{1}{\sqrt{2}} e_1 \right).
\end{align}

One might hope to apply the WGC despite a complete breaking of the SO(3) gauge symmetry, if some of the gauge fields have masses below the Hubble scale and therefore still effectively mediate long-range forces within a Hubble patch, however, in this model this is not the case. 
The masses of the gauge bosons can be determined from the eigenvalues of 
\begin{align}
m^2_{AB} =\frac{1}{2}\big( \mathcal{E}^\Lambda_{\ A} k_\Lambda^\alpha g_{\alpha \bar{\beta} } k_\Sigma^{\bar{\beta}} \mathcal{E}^\Sigma_{\ B} + h.c. \big),
\end{align}
which are $(0 \ ,\ 8 e_1^2 \ ,\ 4 e_1^2 \ , \ 4 e_1^2 )$. The zero mass corresponds to the unbroken U(1) under which the gravitino is uncharged. The remaining masses are clearly of order the Hubble scale and thus do not mediate long-range forces.

\subsection{SO(3) $\times$ O(1,1)} \label{so3o11}

A very instructive model is the one described by (\ref{scalar2man}), but now with a SO(3) $\times$ O(1,1) gauge group.
This model also has a critical point at (\ref{critcpoint}), with a vacuum energy
\begin{align}
{\cal V} = 2\, e_0^2 + 3\, e_1^2 \, , 
\end{align}
and scalar mass spectrum 
\begin{align}
	\begin{split}
		m^2_{(multiplicity)} &= \left( 0_{(1)} \ , \ 2(e_0^2 - e_1^2)_{(3)} \ , \ 4 e^2_{0 \ (1)} \ , \ 4 e^2_{1 \ (2)} \ , x_{1 \ {(1)}} \ , \ x_{2 \ {(1)}} \ , \ x_{3 \ {(1)}} \right. \\[2mm]
		&\quad\quad \left. e_1^2 + \sqrt{4 e_0^4 - 4 e_0^2 e_1^2 + 9 e_1^4}_{(2)} \ , \ e_1^2 - \sqrt{4 e_0^4 - 4 e_0^2 e_1^2 + 9 e_1^4}_{(2)} \right) \times {\cal V} \, , 
	\end{split}
\end{align}
where $x_{1,2,3}$ are solutions of the cubic equation
\begin{align}
x^3 + 6 e_1 x^2 + \left(4 e_0^2 e_1^2 - 4 e_0^4\right) \,x - \left(16 e_0^4 e_1^2 - 16 e_0^2 e_1^4 + 32 e_1^6\right) = 0 \ .
\end{align}
This mass spectrum pushes the limits of the de Sitter criterion, but not parametrically so. 
Again, the normalization of the vector kinetic terms, given by  the values of the gauge kinetic functions at the critical point, is trivial, because ${\cal I}_{\Lambda \Sigma} = -\frac12 \, \delta_{\Lambda \Sigma}$.

As in the previous model, the gravitino masses vanish at the central point, but, once again, 
the unbroken SO(3) prevents a straightforward application of the WGC 
and one must therefore resort to the argument presented in section \ref{non-abelian} 
where we look at nearby configurations that break the SO(3) symmetry down to a U(1).  
In particular, for the special choice $e_0 = e_1$, we find entire lines of critical points that pass from the center and 
are parametrized by any of the three imaginary components. 
Along each of these lines, except the central point, the SO(3) is broken to a U(1) and therefore we can confidently invoke the WGC.

For the rest of this subsection we will set $e_0=e_1$ and for concreteness let us take $\text{Im}\, z_2=z$ as the modulus, with all other scalars remaining fixed at zero. 
The scalar mass spectrum along this line is given by
\begin{align}
	\begin{split}
		m^2_{(multiplicity)}&=\left(0_{(4)},-2/5_{(1)}, 4/5_{(3)},\frac{4}{5-5z^2}_{(1)}, x_{1 \ (1)} , x_{2 \ (1)} , x_{3 \ (1)} \right. \\[2mm] 
		& \quad\quad \left.\frac{1+z^2+\sqrt{9 - 14 z + 9 z^2}}{5 - 5z^2}_{(1)},\frac{1+z^2-\sqrt{9 - 14 z + 9 z^2}}{5 - 5z^2}_{(1)}\right) \times {\cal V},
	\end{split}
\end{align}
where $x_{1,2,3}$ are now solutions of
\begin{align}
	\begin{split}
		& x^3 + \sqrt{1 - z^2}(6 - 
  2 z^2 ) x^2 + (16 z^2 - 32 z^4 + 
  16 z^6) x  \\[2mm]
	 & + \sqrt{1 - z^2}(-32 + 128 z^2 - 
 192 z^4 + 128 z^6 - 
 32 z^8 ) = 0 \, . 
 	\end{split}
\end{align}
\begin{figure} 
\includegraphics[scale=0.6]{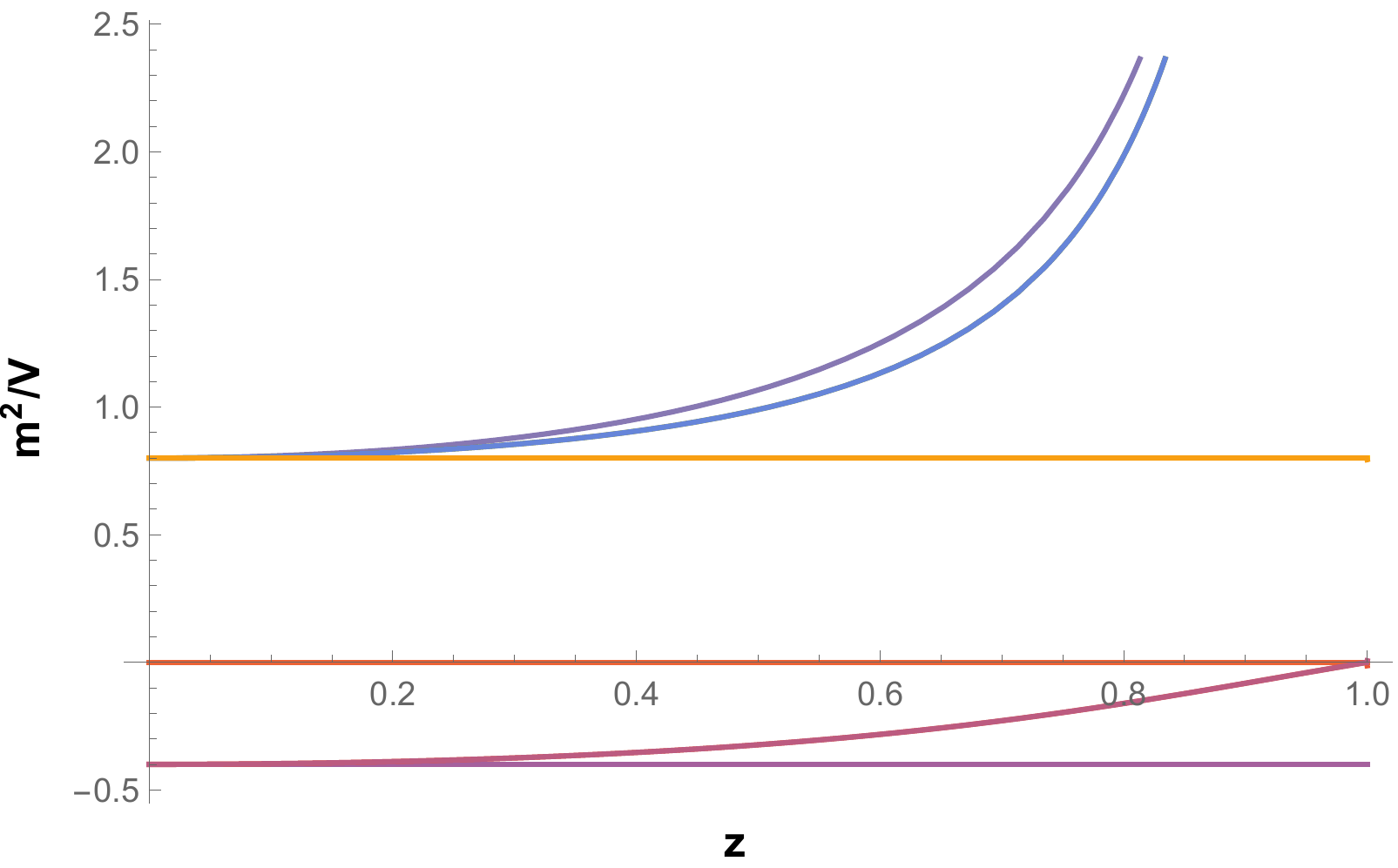} 
\centering
\caption{The ratio $m^2/{\cal V}$ of all the scalar fields in the SU(2)$\times$O(1,1) model, plotted as a function of the modulus $z = \text{Im}\, z_2$. The central vacuum $z=0$ preserves the full SU(2) gauge group. Tachyons with $m^2/{\cal V} = -2/5$ are present for all values of $z$. }
\label{massplot}
\end{figure}
The behavior of these normalized masses is shown in figure \ref{massplot}. 
Notice that, in addition to the three goldstone modes of the SU(2)$\times$O(1,1) breaking to U(1), there is an additional massless scalar field, $z = \text{Im} z_2$.
We therefore now have a residual U(1) gauge group with respect to which the gravitini are charged.
This is specified by the physical gravitino charges 
\begin{align}\label{chargesgrav23}
q_{A} = \left( \pm 0 \ , \ \pm \ e_1 \ , \ \pm \ e_1 \sqrt{\frac{1+z^2}{1-z^2} } \ , \ \pm \ e_1 \right),
\end{align}
where the normalizations follow from the gauge couplings at the line of critical points
\begin{align}
-\frac{1}{2}\,\mathcal{I}^{-1} = \begin{pmatrix} \frac{1+z^2}{1-z^2} & 0 & 0 & 0 \\
											0 & 1 & 0 & 0 \\
											0 & 0 & \frac{1+z^2}{1-z^2} & 0 \\
											0 & 0 & 0 & 1 \end{pmatrix}.
\end{align}
The third eigenvalue of (\ref{chargesgrav23}) is the coupling of the unbroken U(1) subgroup of the SO(3) symmetry and thus is the one that can be used for the WGC. 
This implies that the WGC cut-off is 
\begin{equation}
		\Lambda_{UV} = e_1\, \sqrt{\frac{1+z^2}{1-z^2}} \, . 
\end{equation}
On the other hand the gravitino mass matrix is
\begin{align}
S_{ij} = \begin{pmatrix} e_1\frac{z}{\sqrt{1-z^2}} & 0 \\ 0 & e_1 \frac{ z}{\sqrt{1-z^2}} \end{pmatrix}
\end{align}
and therefore, as we move away from the central point, both gravitini become massive, while the gauge coupling increases, thus increasing the magnetic WGC cut-off. 

\begin{figure} 
\includegraphics[scale=.7]{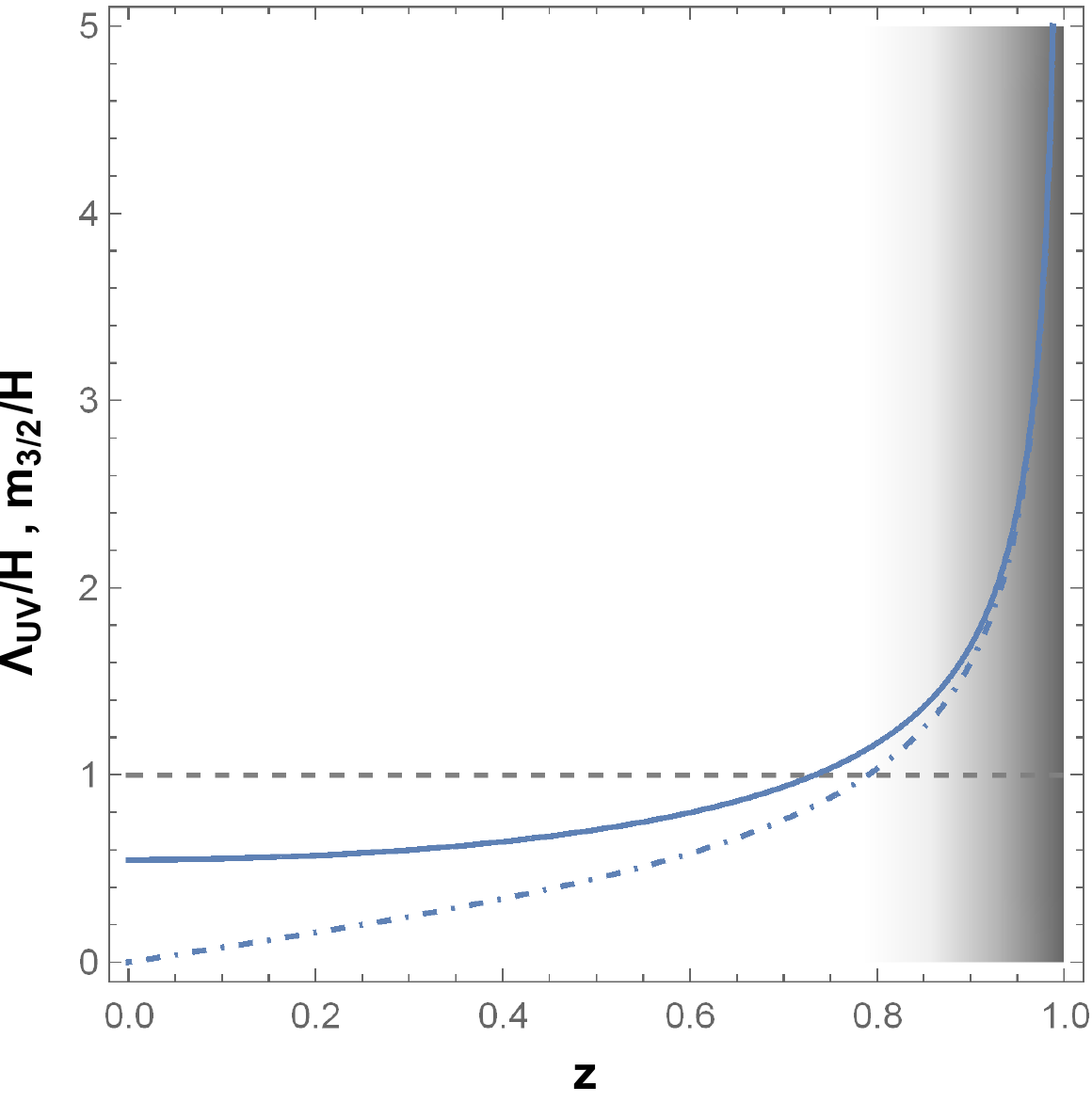} 
\centering
\caption{The ratios $\Lambda_{UV}/H$ (solid) and $m_{3/2}/H$ (dot-dashed) as a function of the modulus $z=\text{Im}z_2$ in the SU(2)$\times$O(1,1) model. For small values of $z$ the gravitino mass vanishes and the Hubble scale is above the cut-off, so the theory has a Dine--Seiberg problem. 
The shaded gray region denotes where the effective theory is increasingly well controlled; in the dark gray part $H \ll \Lambda_{UV}$. The gravitino mass is always below the cut-off, approaching it as $z$ approaches the boundary of moduli space.}
\label{cut-offplot}
\end{figure}

As we mentioned before, this is very instructive for various reasons.
First of all, we see that the vacuum at $z=0$ has a symmetry enhancement, so we could not apply directly the WGC.
However, this vacuum is now the limiting point of a series of critical points with a residual U(1) gauge group for which we can apply the WGC and having charged gravitini, we can apply our argument.
In fact, all the critical points close to the central one are not part of a good EFT, because the vacuum energy is larger than the cut-off scale ($\Lambda_{UV}/H < 1$).
Then, as the gravitini mass approaches the Hubble, the Hubble parameter itself becomes smaller than the cut-off energy, saving the EFT approximation.
This can be seen explicitly in figure \ref{cut-offplot}, where both the ratios of the cut-off over Hubble and gravitini mass over Hubble are plotted, for the whole range of validity of the scalar vacuum expectation value $z$.

\section{Unstable vacua with no massless U(1) couplings} 
\label{nou1}

We conclude our discussion of N=2 models by giving some simple models for which our criteria do not apply, just to clarify the existence of situations that avoid our assumptions.

\subsection{Massive gravitini} 

The first simple example is based on a method to construct stable de Sitter critical points in N=2 presented in \cite{Catino:2013}. 
Such critical points always have non-vanishing gravitini mass and therefore our general argument does not apply. 
In addition as the only gauging involved is a Higgsed U(1) the WGC cannot be applied to eliminate such vacua. 
Taking however into account that in \cite{Catino:2013} only a general strategy is presented but no explicit example is given, we believe it is useful to see a concrete example.
We thus skip the general properties presented in \cite{Catino:2013} and focus on a model that contains a single hypermultiplet with a Quaternionic-K\"ahler geometry that is not homogeneous.
This is a special case of the general metric presented in \cite{Catino:2013} of a Quaternionic-K\"ahler with one isometry.
The specific example we study has a metric given by
\begin{align}
ds^2 = \frac{1}{2 \rho^2} 
\left(f d\rho^2+ f e^h(d\theta^2+d\tau^2)
+ f^{-1} (d\sigma+\Theta)^2 
\right),
\end{align} 
where the scalar fields are $q^u = \{\rho, \theta, \tau, \sigma\}$ and 
\begin{align}
h = \log ( a \rho + b ) \ , \quad f = \frac{a \rho + 2 b}{a \rho + b} 
\ , \quad \Theta = \frac{a}{2} \left( \theta d \tau - \tau d \theta \right) \, ,
\end{align} 
for $a,b$ real parameters.
This metric has an obvious shift symmetry along $\sigma$.
After the gauging with the graviphoton of such isometry 
\begin{align}
k^u = e_0 \, (0,0,0,1) \, , 
\end{align}
we have a scalar potential of the form 
\begin{align}
\mathcal{V} = \frac{e_0^2}{\rho^2} \left( \frac{a \rho + b}{a \rho + 2b} - \frac34 \right) \,. 
\end{align}
This has a critical point at 
\begin{align}
\rho = \frac{(1 + \sqrt{5}) b}{a} \,, 
\end{align}
which gives a positive vacuum energy 
\begin{align}
{\cal V} = \frac{(5 \sqrt 5 - 11) a^2 e_0^2}{32 b^2} \,. 
\end{align} 
The scalar $\rho$ is tachyonic of course, 
while the scalars $\theta$ and $\tau$ are flat directions. 
In particular the canonically normalized mass of $\rho$ is given by 
\begin{align}
m_\rho^2 = - (5 + \sqrt{5} )\times {\cal V} \,.
\end{align}
As a result the refined de Sitter conjecture \cite{Andriot:2018wzk,Garg:2018reu} is not violated. 
We can also compute the gravitini masses 
\begin{align}
m_{3/2}^2 = \frac{e_0^2 a^2}{(2 (1 + \sqrt{5}) b)^2} \,. 
\end{align}

We see that this critical point is not threatened by a low cut-off because there is no U(1) to invoke the WGC, 
and at the same time both gravitini are massive. 
One can work out different examples that include additional vector multiplets, but they essentially have the same property as far as our work is concerned. 
It is further worth noting that in the procedure of \cite{Catino:2013} for constructing fully stable de Sitter vacua,
there is always a spectator U(1) related to the graviphoton because the shift symmetry of the hypermultiplet is gauged 
by a vector belonging to a physical vector multiplet. 

\subsection{Massless uncharged gravitini} 
\label{MUG}

Another simple instance where our argument in section \ref{general} does not apply is when the gravitini are both massless and uncharged. 
A model of this type is simply obtained from the models of subsection \ref{so3o11} by taking $e_1 = 0$, i.e. removing the SO(3) gauging. 
The central vacuum is still present, with energy ${\cal V}= 2e_0^2$, but there is no longer any physical U(1) charge to determine a WGC cut-off. 
Indeed, the vacuum energy only comes from the charges of the hypermultiplets under the broken symmetries, in this case our O(1,1). 
Note that there are three more vectors in the theory on top of the graviphoton, but they are merely spectators and one can truncate them without changing the properties of the example. 

We see that the fact that we cannot exclude the existence of this de Sitter vacuum due to the massless but uncharged gravitini aligns nicely with the FL bound \cite{Montero:2019ekk,Montero:2020rpl,Montero:2021otb} which does not prohibit massless uncharged fields either. 
Furthermore, models that contain only hypers and where the gauging is from the graviphoton have been proven to be tachyonic \cite{Gomez-Reino:2008ylr}. 
In particular as shown in \cite{Gomez-Reino:2008ylr}, such models always contain an $\mathcal{O}(1)$ in Hubble units mass tachyon and so do not violate the de Sitter criterion \cite{Obied:2018sgi,Andriot:2018wzk,Garg:2018reu}.

\section{Maximal supergravity with light gravitini} 
\label{sec:massless_gravitini_in_maximal_supergravity}

While all the discussions and examples provided so far are related to N=2 gauged supergravity, we strongly believe that our arguments are fairly general and should apply to any gauged extended supergravity theory.
In this section, we show how the proof of section \ref{general} can be applied to the case of maximal supergravity.
We believe that similar results could be obtained for any N $\geq 2$.

Let us first recall some crucial aspects of N=8 gauged supergravity \cite{deWit:2007kvg}.
Maximal supergravity contains a single gravity multiplet, whose fields are the graviton, $g_{\mu\nu}$, 8 gravitini, $\psi_\mu^i$ ($i = 1,\ldots,8$), 28 vector fields, $A_\mu^\Lambda$ (conventionally $\Lambda=0,\ldots,27$), 56 spin 1/2 dilatini, $\chi_{ijk} = \chi_{[ijk]}$, and 70 real scalar fields, $\varphi^u$ ($u = 1, \ldots,70$). 
The scalar fields describe a non-linear $\sigma$-model given by a homogeneous manifold 
\begin{equation}
		{\cal M}_{\textrm{scalar}} = \frac{{\rm E}_{7(7)}}{{\rm SU}(8)}.
\end{equation}
The vector fields and their duals transform in the \textbf{56} dimensional fundamental representation of E$_{7(7)}$, which is a symplectic representation, defining an embedding of E$_{7(7)}$ in Sp$(56,{\mathbb R})$, i.e.~$V^M = \{V^\Lambda, V_{\Lambda}\}$.
The coset representative is customarily described by complex 56-dimensional vectors, $L_M{}^{ij} = -L_M{}^{ji}$, and their complex conjugates, $L_{M\,ij}$, which together build a matrix
\begin{equation}
		L_M{}^{\underline{N}} = \left(L_M{}^{ij}, L_{M\,kl}\right).
\end{equation}
This matrix transforms under rigid E$_{7(7)}$ transformations from the left and under local SU(8) transformations from the right.
We also note the following properties of $L_M{}^{\underline{N}}$, which follow from their definition, 
\begin{equation}\label{cosetrelations}
	\begin{split}
	L_M{}^{ij} L_{N\,ij} - L_N{}^{ij} L_{M\,ij} &= i\, \Omega_{MN}, \\[2mm]
	\Omega^{MN} L_M{}^{ij} L_{N\,kl} &= i\, \delta^{ij}_{kl}, \\[2mm]
	\Omega^{MN} L_M{}^{ij} L_{N}^{kl} &= 0,
	\end{split}
\end{equation}
where $\Omega$ is the symplectic invariant matrix. 

The gauging procedure fixes the gauge generators $X_M$ from the E$_{7(7)}$ ones $t_{\alpha}$ specifying the embedding tensor $\Theta$
\begin{equation}
		X_{MN}{}^P = \Theta_M{}^\alpha (t_{\alpha})_N{}^P.
\end{equation}
Of course, consistent gaugings have restrictions on the allowed form of $\Theta$ and consequently of $X$, as discussed in \cite{deWit:2007kvg}.

We are interested in the vector kinetic terms, in the Lagrangian sector describing the kinetic and mass terms for the gravitini, and in the scalar potential. 
The vector kinetic terms have the usual form 
\begin{align}
e^{-1} {\cal L}_{kin.} = \frac14\, {\cal I}_{\Lambda \Sigma} \, F_{\mu \nu}^\Lambda F^{\mu \nu\Sigma} \, ,
\end{align}
although now the vector kinetic term can be expressed in terms of the coset representatives as
\begin{equation}
  \label{eq:inv-Im-N}
  \mathcal{I}^{-1|\Lambda\Sigma}
  = -2\,L^\Lambda{}_{ij} \,L^{\Sigma\,ij}\,.
\end{equation}

The relevant sector for the gravitini is
\begin{equation}
		-\frac 12\varepsilon^{\mu\nu\rho\sigma}\,
		   \left(\bar\psi_{\mu}{}^{i}\gamma_{\nu}{D}_{\rho}(\omega,{\cal Q})\psi_{\sigma\,i} + h.c.\right) +g\, e\left(\frac12\sqrt{2}\,  A_{1\,ij}\,
\bar{\psi}\,^i_\mu \gamma^{\mu\nu} \psi^j_\nu + h.c.\right),
\end{equation}
where the covariant derivative acts as 
\begin{equation}
		{D}_{\rho}(\omega,{\cal Q})\psi_{\sigma\,i} = {D}_{\rho}(\omega)\psi_{\sigma\,i} 
		+ \frac12 {\cal Q}_{\rho\,i}{}^j \psi_{\sigma\,j} \, , 
\end{equation}
and the gauged SU(8) connection ${\cal Q}_\mu$ contains the gauging charges ${\cal Q}_{Mi}{}^j$ in the form 
\begin{equation}
  \label{eq:expression-Q}
  {\cal Q}_{\mu \,i}{}^{j} =  \frac23 \mathrm{i} ( L_{\Lambda\,ik}
  \,\partial_\mu L^{\Lambda\,jk} - L^\Lambda{}_{ik}
  \,\partial_\mu L_\Lambda{}^{\!jk})    - g\, A_\mu{}^M\, 
  {\cal Q}_{M\,i}{}^{j}   \,.
\end{equation}
The explicit expression of the gauging charges can be obtained from the following identities,
\begin{equation}
  \label{eq:P-Q-gauged}
  \begin{split}
{\cal Q}_{M\, ij}{}^{\!kl} &= \delta_{[i}{}^{[k}\, {\cal
  Q}_{M\,j]}{}^{\!l]}\,  = i \,\Omega^{NP} \,L_{N\,ij} X_{MP}{}^Q \,L_Q{}^{kl}\;,\\[2mm]
{\cal P}_{M\, ijkl} & = \frac1{24}\,\varepsilon_{ijklmnpq}\, {\cal P}_{M}^{mnpq} = i\, \Omega^{NP} \,L_{N\,ij} X_{MP}{}^Q \,L_{Q kl}{} \;,
   \end{split}
\end{equation}
where ${\cal Q}_{M}{}^{\!i}{}_{\!j} = - {\cal Q}_{M j}{}^i$ and ${\cal Q}_{M i}{}^i=0$, which means that ${\cal Q}_{M i}{}^j$ is taken to be antihermitian ${\cal Q}_M^\dagger = - {\cal Q}_M$. 
The Lagrangian mass $A_{1\,ij}$ is defined by the gauging procedure, together with the tensor $A_{2i}{}^{\!jkl}$, which will fix the scalar potential, as 
\begin{equation}
	 \mathrm{i}\,\Omega^{MN}  \,{\cal Q}_{M\,i}{}^j\,L_N{}^{kl} = - \,A_{2i}{}^{jkl} - 2\, A_1{}^{j[k}\,\delta^{l]}{}_i \,.   
\end{equation}
Finally, the scalar potential can be written by using the various structures we introduced so far as
\begin{eqnarray}
  \label{eq:g-potential}
  \mathcal{V}&=& g^2 \Big \{\frac{1}{24}  \left| A_{2i}{}^{\!jkl}\right|^2 -
  \frac{3}{4} \left| A_1^{ij}\right|^2\Big\} \nonumber\\[1ex]
  &=& \frac{1}{336}\, g^2 \,\mathcal{M}^{MN} \left\{ 8\,
  \mathcal{P}_M{}^{ijkl} \mathcal{P}_{N ijkl} 
  + 9\, \mathcal{Q}_{Mi}{}^j\, \mathcal{Q}_{Nj}{}^i  \right\},  
\end{eqnarray}
where
\begin{equation}
  \label{eq:def-M}
  {\cal M}_{MN} \equiv
  L_M{}^{ij} \,L_{N\,ij} + L_{M\,ij}\, L_N{}^{ij}\, , \qquad   {\cal M}^{MN}= \Omega^{MP} \Omega^{NQ}{\cal M}_{PQ} \,,
\end{equation}
and one notes the relations
\begin{eqnarray}
  \label{eq:XX-PP-QQ}
  \mathcal{M}^{MN} \mathcal{P}_M{}^{ijkl} \mathcal{P}_{N ijkl} &=& 
  4\, \vert A_{2l}{}^{ijk}\vert^2  \,,
  \\[2mm] 
  \mathcal{M}^{MN} \mathcal{Q}_{Mi}{}^j\, \mathcal{Q}_{Nj}{}^i &=& 
  -2 \, \vert A_{2l}{}^{ijk}\vert^2 -28\, \vert A_1{}^{ij}\vert^2 \,.
\end{eqnarray}

Now that all necessary ingredients have been put forward, we can build our argument along the lines of section \ref{general}.

First of all we assume that all gravitini are massless and hence that we have a de Sitter critical point where $A_{1}^{ij} = 0$.
This implies that at the critical point it holds that 
\begin{equation}
		\mathcal{M}^{MN} \mathcal{P}_M{}^{ijkl} \mathcal{P}_{N ijkl} = -2 \mathcal{M}^{MN} \mathcal{Q}_{Mi}{}^j\, \mathcal{Q}_{Nj}{}^i
\end{equation}
and in turn that the potential can be written as
\begin{equation}
	    \mathcal{V}= -\frac{1}{48}\, g^2 \,\mathcal{M}^{MN}  \mathcal{Q}_{Mi}{}^j\, \mathcal{Q}_{Nj}{}^i > 0\,.
\end{equation}
If we move to an electric symplectic frame, we can further simplify this to
\begin{equation}
	    \mathcal{V}= -\frac{1}{48}\, g^2 \,\mathcal{M}^{\Lambda \Sigma}  \mathcal{Q}_{\Lambda i}{}^j\, \mathcal{Q}_{\Sigma j}{}^i > 0\,
\end{equation}
and using (\ref{eq:inv-Im-N}), (\ref{eq:def-M}) and the coset relations (\ref{cosetrelations}) we find ${\cal M}^{\Lambda \Sigma} = -{\cal I}^{\Lambda \Sigma}$
\begin{equation}
		\mathcal{V}= \frac{1}{48}\, g^2 \,\mathcal{I}^{\Lambda \Sigma}  \mathcal{Q}_{\Lambda i}{}^j\, \mathcal{Q}_{\Sigma j}{}^i = -\frac{1}{48}\, g^2 \,\mathcal{I}^{\Lambda \Sigma}  Tr({\cal Q}_\Lambda^\dagger {\cal Q}_\Sigma) > 0 \, , 
\end{equation}
where we recall that $\mathcal{I}$ is negative definite and ${\cal Q}$ is anti-hermitian.

At this point the argument follows along the same lines as in section \ref{general}.
We define a set of vielbeins to put the kinetic terms of the vectors in canonical form
\begin{align}
- {\cal I}_{\Lambda \Sigma} = \delta_{AB}\, {\cal E}^A_\Lambda {\cal E}^B_\Sigma \ , \qquad \quad {\cal E}^A_\Lambda {\cal E}^\Lambda_B = \delta_B^A \, , 
\end{align}
use the same vielbeins to identify the physical, now hermitian, charges of the gravitini
\begin{equation}
Q_A = \frac{i}{2}\, g\, {\cal E}_A^\Lambda \, {\cal Q}_\Lambda \, ,
\end{equation} 
so that (6.8) becomes
\begin{equation}
        D_\rho \psi_{\sigma i} = D_{\rho}(\omega)\psi_{\sigma i} + \ldots + i A_\mu^A Q_{A\,i}{}^j \psi_{\sigma j},
\end{equation}
and the scalar potential at the critical point is
\begin{equation}
        {\cal V} = \frac{1}{12}\,\delta^{AB}\, {\rm Tr}\left( Q_A Q_B \right) \,. 
\end{equation}  
We therefore see that if there is a U(1) surviving at the critical point under which the gravitini are charged, the scalar potential is larger than the sum of its square.

Clearly the only effect of switching-on a parametrically small gravitino mass is to slightly alter the vacuum energy. 
However, as long as such contribution is parametrically smaller than the Hubble scale it does not alter the fact that the vacuum energy hits the cut-off. 
As a result we conclude once again that (quasi) de Sitter with light charged gravitini belongs to the swampland. 

So far we do not have many examples of de Sitter vacua in maximal gauged supergravity and all the examples we have do not have abelian factors in the residual gauge symmetry.
However, we can once more employ the argument about the WGC for critical points with a non-abelian gauge symmetry made in section \ref{general}, because in maximal supergravity every time we gauge a non-abelian symmetry of the scalar manifold, we will have scalars that potentially break this symmetry by acquiring a vev.
This would tell us that the SO(4) $\times$ SO(4) vacuum of the SO(4,4) gauging in \cite{DallAgata:2012plb} is not a consistent effective theory, because it has massless gravitini, charged under the residual gauge group.
While this was expected for the vacuum coming from the regular gauging, which is a consistent truncation of type II compactifications on a hyperboloid, and therefore do not have a mass gap with the Kaluza--Klein states, it is certainly interesting for the deformed models, where the SO(4,4) gauge group was embedded in a new, rotated way inside E$_{7(7)}$.
We also confirm that the second de Sitter vacuum found in \cite{DallAgata:2012plb}, when the deformation parameter is non-vanishing, has non-vanishing gravitino masses and therefore could still survive the WGC constraints, while having a parametrically small tachyon.\footnote{
See also \cite{Andriot:2021rdy} for further solutions of Type II on non-compact group manifolds.}


\section{Conclusions} \label{concl}

In this paper, we have argued that de Sitter critical points in extended supergravity violate the magnetic weak gravity conjecture when they have charged, light gravitini. 
We have presented a general proof of this claim in N=2 and N=8 gauged supergravity. 
We have further illustrated this claim with several N=2 models with hypermultiplets, whose scalar potentials admit de Sitter critical points, both stable and unstable. 
We also presented examples of critical points that escape the WGC by having either massive gravitini or no U(1) gauge symmetry at the critical point, and a model where a modulus interpolates between respecting and violating the WGC. 
Many of the unstable critical points that we rule out respect the de Sitter criterion. 
Interestingly, our results are especially consonant with the festina lente bound, which forbids charged particles that are too light in a de Sitter background. 
Our findings are also similar in spirit to other works pointing towards a lowering of the UV cut-off of effective theories in the limit of vanishing gravitino mass.

\section*{Acknowledgements}

\noindent
GD is supported in part by MIUR-PRIN contract 2017CC72MK003, FF and ME are supported by the STARS grant SUGRA-MAX.

\appendix

\section{Basic Ingredients of 4D N=2 gauged supergravity} \label{conventions}

In this appendix we give a very short description of some identities in Special-K\"ahler geometry, Quaternionic-K\"ahler geometry and in the gauging procedure of N=2 supergravity.
It is by no means a comprehensive summary of gauged N=2 supergravity theories, but it contains the ingredients required to reproduce the calculations in this work.
For the derivation of what follows we refer the reader to the original works we used \cite{Strominger:1990pd,Ceresole:1995jg,Andrianopoli:1996cm,Ceresole:1995ca,DAuria:2001rlt} and to the very nice review \cite{Lauria:2020rhc}.

We should start by stressing that although a full N=2 duality covariant supergravity action has not been built so far, decisive steps have been taken in this direction.
As shown in \cite{DallAgata:2003sjo}, whenever one introduces magnetic gaugings, tensor multiplets have to be introduced. 
In the case of supergravity coupled to vector multiplets, one has therefore to improve couplings to vector-tensor multiplets \cite{DAuria:2004yjt,Andrianopoli:2007ep} (and its extension to non-trivial FI terms in \cite{DallAgata:2010ejj}). 
For the general matter-coupled case, an outline of the general procedure by using the embedding tensor formalism can be found in \cite{Louis:2009xd} and general Lagrangians for N=2 conformal supergravity theories with arbitrary gaugings have been presented in \cite{deWit:2011gk}.
Our formulae are straightforward applications of the results contained in the above references.

\subsection{Vector multiplets}

The geometry described by the scalar fields appearing in N=2 vector multiplets coupled to supergravity is called Special-K\"ahler.
A Special-K\"ahler manifold is parameterized by complex coordinates $z^I$, $I=1,\dots,n_V$.
Since this is the geometry of the vector-multiplet sector, electric-magnetic duality plays a role in constraining the manifold and this is made manifest by describing the geometry by means of holomorphic sections
\begin{equation}
Z^M = \begin{pmatrix} X^\Lambda(z) \\ F_\Lambda(z) \end{pmatrix} \ , \qquad \Lambda = 0, I \, ,
\end{equation} 
where the additional sections with index 0 have been added to take into account the graviphoton and its dual, which do not have corresponding scalars in their multiplet.
When a prepotential $F(X)$ exists, these sections can also be though of as projective coordinates and $F_{\Lambda} = \partial_{\Lambda} F(X)$.
However, special geometry can be defined in the absence of such a prepotential and, unless specified otherwise, we will not assume the sections are chosen in such a specific frame.
Note that two different patches of the manifold are related by
\begin{equation}\label{patches}
	Z'(z) = e^{-f(z)} S Z(z),
\end{equation}
where $S$ is a constant symplectic matrix and $f$ is a holomorphic function of the coordinates, generating the K\"ahler transformations of the K\"ahler potential.
Defining the symplectic product
\begin{equation}
\langle A, B\rangle = A^T \Omega B =A^\Lambda B_\Lambda - B^\Lambda A_\Lambda,
\end{equation}
the K\"ahler potential is then
\begin{equation}
K = - \log\left[ -i \langle Z, \bar Z \rangle\right]
\end{equation}
and changing patches, from (\ref{patches}), we get the usual K\"ahler transformation
\begin{equation}\label{kaehler}
	K'(z,\bar{z}) \to K(z,\bar{z}) + f(z) + \bar{f}(\bar{z}).
\end{equation}
On the Hodge bundle over the manifold one can also define covariantly-holomorphic sections
\begin{equation}
V^M = e^\frac{K}{2}Z^M
\end{equation}
such that the whole geometric structure gets encoded in the following algebraic and differential constraints:
\begin{align}
\langle V, \overline V \rangle &= i,\\
U_I &= D_I V = (f_I^\Lambda, h_{I \Lambda}),\\
D_I U_J &= i \, \hat C_{IJK}\,g^{K\bar K}\overline U_{\bar K},\\
D_I \overline U_{\bar J} &= g_{I \bar J}\,\overline V,\\
D_I \overline V &=0,
\end{align}
where now $D_I$ is the covariant derivative with respect to the usual Levi--Civita connection and the K\"ahler connection $\partial_I K$. 
This means that under a K\"ahler transformation \eqref{kaehler}, a generic field $\chi^I$, with charge $p$, namely transforming as $\chi^I \to e^{-\frac{p}{2}f+\frac{\bar{p}}{2}\bar{f}} \chi^I$, has covariant derivative
\begin{equation}
	D_I \chi^J = \partial_I \chi^J + \Gamma_{JK}^I \chi^K +\frac{p}{2} \, \partial_J K \, \chi^I,
\end{equation}
and analogously for $\overline{D}_{\bar{J}}$, with $p \to \bar{p}$.
We followed standard conventions and chose $p = -\bar{p} = 1$ for the weight of $V$.
Note also that
\begin{equation}
	g_{I\bar J} = i\, \langle U_I , \overline{U}_{\bar J}\rangle.
\end{equation}

One more ingredient needed is the matrix defining the non-minimal couplings of the vector multiplets
\begin{equation}
	{\cal N}_{\Lambda \Sigma} = {\cal R}_{\Lambda \Sigma} + i\, {\cal I}_{\Lambda \Sigma} = \left(M_{\Lambda}, \overline{h}_{\bar{I}}\right) \left(L^\Sigma, \overline{f}_{\bar{I}}^\Sigma\right)^{-1},
\end{equation} 
which results in the kinetic Lagrangian for the vector multiplets 
\begin{equation}
	{\cal L}_{kin} = \frac14 \,e\, {\cal I}_{\Lambda \Sigma} \,F^\Lambda_{\mu\nu} F^{\Sigma \mu\nu} 
	+ \frac18 \, {\cal R}_{\Lambda \Sigma}\, \epsilon^{\mu\nu\rho\sigma} \,F_{\mu\nu}^\Lambda F^\Sigma_{\rho\sigma},
\end{equation}
which means that ${\cal I}$ is negative definite.

The scalar potential following from the gauging procedure has two main contributions. 
The first one ${\cal V}_F$ is coming from the $N=2$ Fayet--Iliopoulos terms, which are the relics of the possible coupling to hypermultiplets.
If we consider full symplectic invariance, the FI terms are given in terms of the triplet of FI charge vectors $Q^{M x} = \left(P^{\Lambda x}, P_{\Lambda}^x\right)$, with $x=1,2,3$:
\begin{equation}\label{Ftermapp}
	{\cal V}_F = g^{I \bar J} \langle Q^x, U_I \rangle \langle Q^x, \overline{U}_{\bar J}\rangle - 3 \langle Q^x, V\rangle \langle Q^x, \overline V\rangle.
\end{equation}

The second contribution is the $D$-term ${\cal V}_D$ generated by the proper gauging of the isometries of the Special--K\"ahler scalar manifold.
Again, trying to be general and maintaining symplectic invariance, for Special-K\"ahler manifolds, the isometries can be derived by looking at their linear action on the sections.
In fact all isometries must preserve (\ref{patches}) and therefore
\begin{equation}
	\delta_P Z^M = (T_P)_N{}^M Z^N - f_P(z) Z^M,
\end{equation}
where $T_P$ is a symplectic matrix (the generator of $S$) satisfying 
\begin{equation}\label{symplcond}
	T_{\Lambda}^T \Omega + \Omega T_{\Lambda} = 0,
\end{equation}
and $f_N(z)$ are compensating holomorphic functions, which are going to be related to how the K\"ahler potential transforms under such isometries.
Using full Sp($2n_V+2$,${\mathbb R}$) indices:
\begin{equation}\label{symplconstraint}
	T_{M[N}{}^Q \Omega_{P]Q} = 0.
\end{equation}
Consistency of the gauging also requires
\begin{equation}\label{repconstraint}
	T_{(MN}{}^Q \Omega_{P)Q} = 0.
\end{equation}
Note that now the position of the index transforming with $S$ is fixed, so that indices $M,N,\ldots$ are lowered and raised with the symplectic matrix. 
Upper indices transform with $S$ and lower indices transform with $S^{-1} = -\Omega S^T \Omega$, so that $V^M W_M = V^M \Omega_{MN} W^N$ is symplectic invariant
\begin{equation}
	V^{M\prime} W_M^\prime = V^{M\prime} \Omega_{MN} W^{N\prime} = V^P S^M{}_P \Omega_{MN} S^N{}_Q W^Q = V^M \Omega_{MN} W^N = V^M W_M.
\end{equation}
The non-linear action on the coordinates can be obtained by means of holomorphic Killing vectors, which can be related to the linear action above in frames where the prepotential exists.
In this case the Killing vectors follow by introducing \emph{normal coordinates} $z^I \equiv X^I/X^0$:
\begin{eqnarray}
	\delta_M z^I &=& \frac{\delta_{M} X^I}{X^0} - \frac{X^I}{X^0} \frac{\delta_{M} X^0}{X^0} = \nonumber \\ 
	&=& \frac{(T_{M} Z)^I}{X^0} - \frac{X^I}{X^0}\frac{(T_{M} Z)^0}{X^0} \equiv k^I_{M}(z).
\end{eqnarray}
At the infinitesimal level
\begin{eqnarray}
	\delta_M V^N &=& - T_{MP}{}^M V^P, \\[2mm]
	\delta_M W_N &=& T_{MN}{}^P W_P.
\end{eqnarray}
Under an isometry the K\"ahler potential transforms as
\begin{equation}
	\delta_{M} K = - e^{K} \, i\, ( \delta_{M}Z^T \Omega \overline{Z} + Z^T \Omega\, \delta_{M}\overline{Z}) = f_M + \overline{f}_M.
\end{equation}

As is customary in supergravity, the gauging procedure is enforced by the introduction of prepotentials (or moment maps) for the gauged isometries.
In this context, the prepotential definition is 
\begin{equation}
	P_M^0 = -i k_M^i \partial_i K +i \, f_M,
\end{equation}
which, in the frame where a prepotential exists, becomes
\begin{equation}
	P_M^0 = e^K\, \overline{Z}^T \Omega T_M Z = e^K\,T_{MN}{}^Q \Omega_{QP} Z^N \overline{Z}^P \,.
\end{equation}
Prepotentials satisfy the constraint
\begin{equation}\label{prepoconst}
	Z^M(z) P_M^0(z,\bar{z}) = 0
\end{equation}
which also implies 
\begin{equation}
	Z^M(z) k_M^I(z) = 0 \,.
\end{equation}
The relations between the prepotentials and the Killing vectors also imply that 
\begin{equation}\label{relation}
	\bar{Z}^M(\bar{z}) k_M^I(z) = i\, g^{I\bar{J}} \overline{U}_{\bar{J}}^M P_{M}^0 \,.
\end{equation}

After the gauging, the resulting scalar potential is therefore
\begin{equation}\label{Dtermapp}
	{\cal V}_{D_1}  = \overline{V}^M k_M^I V^N \bar{k}_N^{\bar{J}} g_{I \bar J} =  g^{I\bar{J}} U_I^M \overline{U}_{\bar{J}}^N \, P_M^0 P_N^0 \,. 
\end{equation}

\subsection{Hypermultiplets}

Hyper-scalars, $q^u, u = 1,...,4n_H$, span a Quaternionic-K\"ahler manifold, namely a $4n_H$-dimensional real manifold endowed with an invertible metric $h_{uv}$ and a triplet of complex structures $(J^x)_u{}^v$, $x=1,2,3$, satisfying the quaternionic algebra
\begin{equation}
		J^x J^y = - \delta^{xy} {\mathbb 1} + \epsilon^{xyz} J^z \ ,
\end{equation}
and with respect to which the metric is hermitian
\begin{equation}
		(J^x)_u{}^w (J^x)_v{}^t h_{wt} = h_{uv}.
\end{equation}
From the complex structures one can introduce a triplet of 2-forms $K^x = h_{uw} (J^x)_v{}^w \, dq^u \wedge dq^v$, which are proportional to the curvatures of an SU(2) bundle with connections $\omega^x$, namely
\begin{equation}
		R^x = d \omega^x +\frac12\, \epsilon^{xyz} \omega^y \wedge \omega^z = -K^x.
\end{equation}
This implies that the quaternionic structures are preserved by the SU(2) connection, i.e.
\begin{equation}
		\nabla K^x = dK^x  + \epsilon^{xyz} \omega^y \wedge K^z = 0.
\end{equation}

This same structure also implies that for each isometry of the manifold, $\delta q^u = \epsilon^M k_M^u$, we can introduce a triplet of moment maps by
\begin{equation}
		2 \,R_{uv}^x\, k_M^v = \partial_u P_M^x + \epsilon^{xyz} \omega_u^y P_M^z.
\end{equation}
and satisfy the consistency condition
\begin{equation}
		R_{uv}^x \, k_M^u k_N^v + \frac12\,\epsilon^{xyz} P_M^y P_N^z = \frac12\, f_{MN}{}^P P_P^z,
\end{equation}
required by gauge invariance of the N=2 action, where $f_{MN}{}^P$ are the structure constants of the gauge algebra.
Using the properties of the SU(2) curvatures, one can also find
\begin{equation}\label{Psoln}
		2n_H\, P^x_M =- (R^x)_u^{\ v} \nabla_v k_M^u \ .
\end{equation}
In the absence of hypermultiplets we can still introduce constant $P^x_M$, which correspond to the FI-terms of the previous subsection.

The gauging of a non-abelian gauge group introduces a new D-term potential
\begin{equation}
		{\cal V}_{D_2} = 4 \,\overline{V}^M k_M^u V^N \bar{k}_N^{v} h_{uv}
\end{equation}
and the F-term potential gets improved from the U(1) charges of the previous section to the full prepotentials $P^{M x} = \left(P^{\Lambda x}, P_{\Lambda}^x\right)$:
\begin{equation}\label{Ftermapp_2}
	{\cal V}_F = g^{I \bar J} \langle P^x, U_I \rangle \langle P^x, \overline{U}_{\bar J}\rangle - 3 \langle P^x, V\rangle \langle P^x, \overline V\rangle.
\end{equation}

\subsection{Potential, Gravitino mass and charge}

Summarizing, the scalar potential of a generic N=2 matter coupled gauged supergravity theory can be written as the sum of three pieces
\begin{align} \label{pot0}
	\mathcal{V} &= \mathcal{V}_{D_1} + \mathcal{V}_{D_2} + \mathcal{V}_F, \\[2mm]
	{\cal V}_{D_1} &= \overline{V}^M k_M^I V^N \bar{k}_N^{\bar{J}} g_{I \bar J}  =  g^{I\bar{J}} U_I^M \overline{U}_{\bar{J}}^N \, P_M^0 P_N^0, \\[2mm]
	{\cal V}_{D_2} &= 4 \,\overline{V}^M k_M^u V^N \bar{k}_N^{v} h_{uv}, \\[2mm]
	{\cal V}_F &= g^{I \bar J} \langle P^x, U_I \rangle \langle P^x, \overline{U}_{\bar J}\rangle - 3 \langle P^x, V\rangle \langle P^x, \overline V\rangle.
\end{align}

{}A general result of consistent gaugings is that we can always rotate the symplectic frame from which we start in the description of the lagrangian so that the couplings and the potentials result from purely electric gaugings \cite{DallAgataxxx}.
This means that once we introduce new sections as in (\ref{patches}) with an appropriate symplectic matrix $S$, we can write the scalar potential above as
\begin{align} \label{pot}
	\mathcal{V} &= \mathcal{V}_{D_1} + \mathcal{V}_{D_2} + \mathcal{V}_F, \\[2mm]
	{\cal V}_{D_1} &= \overline{L}^\Lambda k_{\Lambda}^I \, L^\Sigma \bar{k}_{\Sigma}^{\bar{J}}\, g_{I \bar J} = U^{\Lambda \Sigma} P_{\Lambda}^0 P_{\Sigma}^0, \\[2mm]
	{\cal V}_{D_2} &= 4 \,\overline{L}^{\Lambda} k_{\Lambda}^u\, V^\Sigma \bar{k}_{\Sigma}^{v}\, h_{uv}, \\[2mm]
	{\cal V}_F &= g^{I \bar J} f_I^\Lambda \overline{f}_{\bar{J}}^\Sigma P_{\Lambda}^x P_{\Sigma}^x - 3 \,L^\Lambda \overline{L}^\Sigma P_{\Lambda}^x P_{\Sigma}^x = \left(U^{\Lambda \Sigma} - 3 L^\Lambda \overline{L}^\Sigma\right) P_{\Lambda}^x P_{\Sigma}^x,
\end{align}
where we note the useful identity
\begin{align} \label{Umatrix}
U^{\Lambda \Sigma}=g^{\alpha \bar{\beta}} f_I^\Lambda f_{\bar{J}}^\Sigma = -\frac{1}{2} \,\mathcal{I}^{-1|\Lambda \Sigma} - \bar{L}^\Lambda L^\Sigma.
\end{align}
Clearly in all these expressions the $L^\Lambda$ refer to the new frame $V'$.

{}From the full Lagrangian \cite{Andrianopoli:1996cm,DAuria:2001rlt,Lauria:2020rhc} we can also extract two ingredients that are central in our analysis, the gravitino mass matrix 
\begin{align} \label{m32}
S_{ij} = \langle P^x , V\rangle\,i\, (\sigma_{x})_i^{\ k} \epsilon_{jk},
\end{align}
and the physical charges of the gravitini, which, in the electric frame, are the eigenvalues of
\begin{align}
(q_A)_i{}^j = \frac{1}{2}\, \mathcal{E}^\Lambda_{A} \, P^x_\Lambda (\sigma^x)_i{}^j
\end{align}
where $\mathcal{E}^\Lambda_A \mathcal{E}^\Sigma_B\, \delta^{AB} = {\cal I}^{-1|\Lambda \Sigma}$. 
These are the charges to be used when determining the magnetic WGC cut-off.

\section{$SO(4,2)/SO(4)\times SO(2)$ coset space hyper geometry} \label{so42}

In this appendix we describve the Quaternionic--K\"ahler geometry of the coset space $\frac{SO(4,2)}{SO(4)\times SO(2)}$, parametrized by the scalars $q^u, \ u = 1,...,8$. 
We start from the $SO(4,2)$ generators
\begin{equation}\label{so42gens}
		(T_{\underline{ab}})_{\underline{c}}{}^{\underline{d}} = \eta_{\underline{c}[\underline{a}} \delta_{\underline{b}]}^{\underline{d}},
\end{equation}
where $\underline{a} = 1,\ldots 6$ is in the fundamental of $\mathfrak{so}(4,2)$ and $\eta_{\underline{ab}} = {\rm{diag}}\{1,1,1,1,-1,-1\}$.

We use $C_1 = T_{\underline{15}}$ and $C_2 = T_{\underline{36}}$ as the non-compact Cartan generators and introduce the following set of positive roots with respect to $C_1$ and $C_2$ 
\begin{equation} \label{posroot}
	\begin{split}
		E_0^{(1,1)} &= \frac{1}{\sqrt{2}} \left(T_{\underline{12}} + T_{\underline{25}} - T_{\underline{16}} + T_{\underline{56}}\right) \ , \quad
		E_0^{(1,-1)} =\frac{1}{\sqrt{2}}\left(T_{\underline{12}} + T_{\underline{25}} + T_{\underline{16}} - T_{\underline{56}}\right) \ , \\ 
		E_a^{(1,0)} &= T_{\underline{13}} + T_{\underline{35}} \ ,  \quad E_b^{(1,0)} = T_{\underline{14}} + T_{\underline{45}} \ , \quad E_a^{(0,1)} = T_{\underline{23}} + T_{\underline{36}} \ , 
		\quad E_b^{(0,1)} = T_{\underline{24}} + T_{\underline{46}} \ , 
	\end{split}
\end{equation}
where the superscripts denote the weights under the non-compact Cartans and the $a,b$ subscript distinguishes between generators of different weight under the remaining compact Cartan. 
Together with $C_1$ and $C_2$, these six generators form a basis for the tangent space of the coset space, which we will collectively denote
\begin{align}
G_a = \big(C_1 \ , \ E_a^{(1,0)} \ , \ E_b^{(1,0)} \ , \ E_0^{(1,-1)} \ , \ C_2 \ , \ E_0^{(1,1)} \ , \ E_a^{(0,1)} \ , \ E_b^{(0,1)} \big) \,. 
\end{align}

We then write the coset representative as
\begin{align}
	\begin{split}
	\mathbb{L} &= \exp \left[ (q_6 - \frac{q_2 q_7}{\sqrt{2}} - \frac{q_3 q_8}{\sqrt{2}}) E_0^{(1,1)} + (q_2 + \frac{q_7 q_4}{\sqrt{2}} ) E_a^{(1,0)} + (q_3 + \frac{q_8 q_4}{\sqrt{2}})E_b^{(1,0)}\right.\\[2mm] 
	& \left.\qquad  + \, q_7 E_a^{(0,1)} + q_8 E_b^{(0,1)} + q_4 E_0^{(1,-1)} \right] \exp \left[\log(q_1) C_1 \right] \exp \left[\log(q_5) C_2 \right] \,, 
	\end{split}
\end{align}
from which we can read off the vielbeins $e^a_{\ m}$ through
\begin{align}
G_a e^a_{\ u} dq^u = \mathbb{L}^{-1} d \mathbb{L} \,. 
\end{align}
The resulting vielbein is
\begin{align} \label{vielb}
e^a_{\ u} = \begin{pmatrix} \frac{1}{\sqrt{2} q_1} & 0 & 0 & 0 & 0 & 0 & 0 & 0 \\
									0 & \frac{1}{\sqrt{2} q_1} & 0 & \frac{q_7}{q_1} & 0 & 0 & 0 & 0 \\
									0 & 0 & \frac{1}{\sqrt{2} q_1} & \frac{q_8}{q_1} & 0 & 0 & 0 & 0 \\
									0 & 0 & 0 & \frac{q_5}{\sqrt{2}q_1} & 0 & 0 & 0 & 0\\
									0 & 0 & 0 & 0 & \frac{1}{\sqrt{2} q_5} & 0 & 0 & 0\\
									0 & -\frac{q_7}{q_1 q_5} & - \frac{q_8}{q_1 q_5} & - \frac{5(q_7^2 + q_8 ^2)}{6 \sqrt{2} q_1 q_5} & 0 & \frac{1}{\sqrt{2} q_1 q_5} & \frac{q_4 q_7}{3\sqrt{2} q_1 q_5} & \frac{q_4 q_8}{3 \sqrt{2} q_1 q_5} \\
									0 & 0 & 0 & 0 & 0 & 0 & \frac{1}{\sqrt{2} q_5} & 0 \\
									0 & 0 & 0 & 0 & 0 & 0 & 0 & \frac{1}{\sqrt{2} q_5}  \end{pmatrix}
\end{align}
and the metric is then 
\begin{eqnarray}
\nonumber 
		ds^2 \!\!&=&\!\!\! h_{uv} dq^u dq^v = \delta_{ab} \,e^a_{ \ u} e^b_{ \ v}  dq^u dq^v 
		\\[1mm]
		\nonumber 
		&=&\!\!\! \frac{1}{q_1^2} \left[ dq_1^2 + q_5^2 dq_4^2 + (dq_2 + \sqrt{2} q_7 dq_4)^2 + (dq_3 + \sqrt{2} q_8 dq_4)^2\right]  
		\\[1mm]
		\nonumber 
		&&\!\!\! + \frac{1}{ 72 q_1^2 q_5^2} \left[6 \sqrt{2} dq_6 - 12 q_7 dq_2 - 12 q_8 dq_3 + 2 \sqrt{2} q_4 (q_7 dq_7 + q_8 dq_8) - 5 \sqrt{2} (q_7^2 + q_8^2) dq_4) \right]^2 
		 \\[1mm]
		&&\!\!\! + \frac{1}{q_5^2} \big(dq_5^2 + dq_7^2 + dq_8^2 \big) 
		\,.  
\end{eqnarray}

The homogeneous nature of the scalar manifold allows us to find the Killing vectors of the SO(4,2) isometries by the action of the generators on the coset representative (see for instance \cite{DallAgataxxx}):
\begin{align} \label{findkills}
k_{T_{\rm{SO}(4,2)}}^u \partial_u \mathbb{L} = T_{\rm{SO}(4,2)} \mathbb{L} - \mathbb{L} \, w^H\, T_{H = \rm{SO}(4) \times \rm{SO(2)}}\,,
\end{align}
where the last term is the H-compensator and cancels the part of the transformation that moves along the coset. 
For each generator this is a set of 15 equations ($\mathbb{L}$ has 15 independent components) in 15 unknowns (8 components for the Killing vector $k^u$ and 7 coefficients for the compensator $w^H$). 
Finding the solution is straightforward, but the resulting expressions are very elaborate and we do not present them here.

Finally, we give here the quaternionic structures
\begin{align}
J^1 = T_{\underline{12}} + T_{\underline{34}} \ , \quad J^2 = -T_{\underline{13}} + T_{\underline{24}} \ , \quad J^3 = T_{\underline{23}} + T_{\underline{14}} \,,
\end{align}
which correspond to a normal $SU(2)$ subgroup of the $SO(4)$. 
The action of the generators on the vielbeins defined in \eqref{vielb} can be deduced from their commutators with the corresponding generators such that
\begin{align}
[J^x , e^a G_a] = (J^x)^a_{ \ b} e^b G_a 
\end{align}
and take the form
\begin{equation}
\begin{split}
J^1 &= \begin{pmatrix} 0 & 0 & 0 & -\frac{1}{\sqrt{2}} & 0 & -\frac{1}{\sqrt{2}} & 0 & 0 \\
							0 & 0 & -1 & 0 & 0 & 0 & 0 & 0 \\
							0 & 1 & 0 & 0 & 0 & 0 & 0 & 0 \\
							\frac{1}{\sqrt{2}} & 0 & 0 & 0 & -\frac{1}{\sqrt{2}} & 0 & 0 & 0 \\
							0 & 0 & 0 & \frac{1}{\sqrt{2}} & 0 & -\frac{1}{\sqrt{2}} & 0 & 0 \\
							\frac{1}{\sqrt{2}} & 0 & 0 & 0 & \frac{1}{\sqrt{2}} & 0 & 0 & 0 \\
							0 & 0 & 0 & 0 & 0 & 0 & 0 & -1 \\
							0 & 0 & 0 & 0 & 0 & 0 & 1 & 0 \end{pmatrix} \ ,  \\							
J^2 &= \begin{pmatrix} 0 & -1 & 0 & 0 & 0 & 0 & 0 & 0 \\
							1 & 0 & 0 & 0 & 0 & 0 & 0 & 0 \\
							0 & 0 & 0 & -\frac{1}{\sqrt{2}} & 0 & -\frac{1}{\sqrt{2}} & 0 & 0 \\
							0 & 0 & \frac{1}{\sqrt{2}} & 0 & 0 & 0 & -\frac{1}{\sqrt{2}} & 0 \\
							0 & 0 & 0 & 0 & 0 & 0 & 0 & 1 \\
							0 & 0 & \frac{1}{\sqrt{2}} & 0 & 0 & 0 & \frac{1}{\sqrt{2}} & 0 \\
							0 & 0 & 0 & \frac{1}{\sqrt{2}} & 0 & -\frac{1}{\sqrt{2}} & 0 & 0 \\
							0 & 0 & 0 & 0 & -1 & 0 & 0 & 0 \end{pmatrix} \ ,  \\							
J^3 &= \begin{pmatrix} 	0 & 0 & -1 & 0 & 0 & 0 & 0 & 0 \\
							0 & 0 & 0 & \frac{1}{\sqrt{2}} & 0 & \frac{1}{\sqrt{2}} & 0 & 0 \\
							1 & 0 & 0 & 0 & 0 & 0 & 0 & 0 \\
							0 & -\frac{1}{\sqrt{2}} & 0 & 0 & 0 & 0 & 0 & -\frac{1}{\sqrt{2}} \\
							0 & 0 & 0 & 0 & 0 & 0 & -1 & 0 \\
							0 & -\frac{1}{\sqrt{2}} & 0 & 0 & 0 & 0 & 0 & \frac{1}{\sqrt{2}} \\
							0 & 0 & 0 & 0 & 1 & 0 & 0 & 0 \\
							0 & 0 & 0 & \frac{1}{\sqrt{2}} & 0 & -\frac{1}{\sqrt{2}} & 0 & 0 \end{pmatrix} \ . 
\end{split}
\end{equation}

\end{document}